\documentclass[twocolumn]{aastex631}
\usepackage{amsmath}
\usepackage{booktabs}
\usepackage{mhchem}

\makeatletter
\newcommand*{\linktocite}[2]{%
  \hyper@natlinkstart{#1}#2\hyper@natlinkend}
\makeatother

\shorttitle{Retrieving Red Edges from Exoplanet Reflection Spectra}
\shortauthors{Gomez Barrientos et al.}

\begin{document}

\title{In Search of the Edge: A Bayesian Exploration of the Detectability of Red Edges in \\ Exoplanet Reflection Spectra}

\correspondingauthor{Jonathan Gomez Barrientos}
\email{jdg276@cornell.edu}

\author[0000-0002-0672-9658]{Jonathan Gomez Barrientos}
\affiliation{Department of Astronomy and Carl Sagan Institute, Cornell University, 122 Sciences Drive, Ithaca, NY 14853, USA}
\affiliation{Division of Geological and Planetary Sciences, California Institute of Technology, Pasadena, CA 91125, USA}
\author[0000-0003-4816-3469]{Ryan J. MacDonald}
\affiliation{Department of Astronomy and Carl Sagan Institute, Cornell University, 122 Sciences Drive, Ithaca, NY 14853, USA}
\affiliation{Department of Astronomy, University of Michigan, Ann Arbor, MI 48109, USA}
\affiliation{NHFP Sagan Fellow}
\author[0000-0002-8507-1304]{Nikole K. Lewis}
\affiliation{Department of Astronomy and Carl Sagan Institute, Cornell University, 122 Sciences Drive, Ithaca, NY 14853, USA}
\author[0000-0002-0436-1802]{Lisa Kaltenegger}
\affiliation{Department of Astronomy and Carl Sagan Institute, Cornell University, 122 Sciences Drive, Ithaca, NY 14853, USA}

\begin{abstract}

Reflection spectroscopy holds great promise for characterizing the atmospheres and surfaces of potentially habitable terrestrial exoplanets. The surface of the modern Earth exhibits a sharp albedo change near 750\,nm caused by vegetation --- the \emph{red edge} --- which would leave a strong spectral signature if present on an exoplanet. However, the retrieval of wavelength-dependent surface properties from reflection spectra has seen relatively little study. Here, we propose a new surface albedo parameterization capable of retrieving the wavelength location of \emph{a priori} unknown `edge-like' features. We demonstrate that a wavelength-dependent surface albedo model achieves higher accuracy in retrieving atmospheric composition. Wavelength-dependent surfaces are also generally preferred over a uniform albedo model when retrieving simulated reflection spectra for a modern Earth analog, even for moderate signal-to-noise ratios ($S/N = 10$) and Earth-like clouds. Further, the location of the modern Earth's red edge can be robustly and precisely constrained (within 70\,nm for $S/N = 10$). Our results suggest that future space-based direct imaging missions have the potential to infer surface compositions for rocky exoplanets, including spectral edges similar to those caused by life on the modern Earth. 

\end{abstract}


\section{Introduction}
\label{sec:intro}

One of the most profound discoveries of the exoplanet era is the realization that rocky planets in the classical habitable zone are prolific around nearby stars \citep[e.g.,][]{Petigura2013,Bryson2021}. Remote observations of such rocky planets hold the potential to deepen our understanding of the physical, chemical, climate, and biological processes shaping these worlds \citep[see e.g.,][for recent reviews]{Kaltenegger2017,Wordsworth2021}. Spectroscopic observations ultimately provide our clearest window into the atmospheric and surface properties of rocky exoplanets.

The near-term focus for rocky exoplanet atmospheric characterization lies with transiting planets orbiting small stars, such as M dwarfs \citep[e.g.][]{Scalo2007,Barstow2016,Morley2017,Fauchez2021} or white dwarfs \linktocite{KalteneggerMacDonald2020}{(e.g., Kaltenegger \& MacDonald et al.} \citeyear{KalteneggerMacDonald2020}, \citealt{Lin2022}). Transmission spectroscopy --- the variation of planetary radius with wavelength --- is currently the most common technique used to diagnose the composition, temperature profile, and aerosol properties of exoplanet atmospheres \citep[e.g.,][]{Brown2001,Sing2016,Kreidberg2018,MacDonald2022}. Existing transmission spectra from the \emph{Hubble Space Telescope} and ground-based facilities have ruled out clear H$_2$-dominated atmospheres for several rocky exoplanets \citep[e.g.,][]{deWit2016,deWit2018,Wakeford2019,Diamond-Lowe2020,Libby-Roberts2021,Garcia2022}. The recently launched JWST will observe terrestrial exoplanets transiting M dwarfs, which should lead to detections of molecules such as CO$_2$, CH$_4$, and H$_2$O \citep[e.g.,][]{Krissansen-Totton2018,Lustig-Yaeger2019,Lin2021}. However, terrestrial exoplanets in the habitable zone of G-type stars like our Sun are not optimal targets for transmission spectroscopy (due to their low planet-star radius ratio, infrequent transits, and atmospheric refraction; see e.g. \citealt{Betremieux2014}).

Directly-detected reflected stellar light is a promising avenue to characterize rocky exoplanets orbiting G-type stars. Detecting reflected light from an Earth-like exoplanet around a star like our Sun requires sensitivity to planet-star contrast ratios at visible wavelengths of $F_p / F_* \sim 10^{-10}$, where $F_p$ and $ F_*$ are, respectively, the fluxes from the planet and star as observed at Earth. The \emph{Roman Space Telescope} (scheduled for launch in 2026) will offer important progress towards this goal, demonstrating space-based coronography by imaging cool giant planets with $F_p / F_* \sim 10^{-8}$ to $10^{-9}$ \citep{Kasdin2020}. Recently, the Astro 2020 Decadal Survey recommended the development of a large IR/Optical/UV space-based telescope --- notionally for launch in the 2040s --- to characterize Earth-like exoplanets around Sun-like stars \citep{NationalAcademiesofSciences2021}. Such a mission would be designed to detect atmospheric and surface biosignatures for a population of Earth-sized exoplanets.

Biosignatures are remotely detectable features indicative of a biological process \citep[for recent reviews, see][]{Kaltenegger2017,Schwieterman2018,Fujii2018}. For exoplanets, biosignatures include the simultaneous detection of a pair of oxidizing and reducing gases (e.g. O$_2$ / O$_3$ and CH$_4$) or temporal variability (e.g, \citealt{Keeling1976}, \citealt{Ford2001,Meadows2006,Meadows2008,Cowan2012,Fujii2017}). Another important class of biosignatures --- and the focus of this study --- are \emph{surface biosignatures}, here defined as remotely detectable spectroscopic features caused by the presence of biology on a planetary surface \citep[e.g.,][]{DesMarais2002,Schwieterman2015a,Hegde2015,Coelho2022}.

The \emph{red edge} is a proposed surface biosignature caused by a biology-induced change in the wavelength-dependent surface albedo. On the modern Earth, where plants cover $\sim$ 60\% of the land area, the red edge can be detected in both resolved reflection spectra \citep{Sagan1993} and disc-averaged Moonshine \citep[e.g.][]{DesMarais2002, Woolf2002,Turnbull2006}. This photosynthetic red edge has changed in strength throughout Earth's history (e.g., from changing surface coverage, types of biota), but should have been detectable for the last 1 billion years \citep{O'Malley-James2018,O'Malley-James2019b}. The origin of Earth's red edge is chlorophyll reflecting more light redwards of $\sim$ 750\,nm. For exoplanets, many studies have suggested that a similar change in the surface albedo at a characteristic wavelength could constitute a biosignature analogous to Earth's red edge \citep[e.g.,][]{Seager2005,Takizawa2017,O'Malley-James2018,O'Malley-James2019b}. Reflected light spectra of an exo-Earth would consequently display a sharp contrast ratio increase if its surface featured a signature like the red edge. This study proposes a method to retrieve the wavelength location of albedo changes, like the red edge, from reflection spectra of rocky exoplanets.

Spectroscopic retrieval is a method commonly used to infer planetary properties (e.g., atmospheric composition, temperature, and clouds) from observed spectra. Bayesian retrieval techniques compare model spectra for a wide range of possible planet properties (typically $\gtrsim 10^5$) to a set of observations, thereby obtaining probability distributions for the planet properties \citep[e.g.,][]{Benneke2012,MacDonald2017,Molliere2019}. Reflected light exoplanet retrieval techniques were initially developed for directly imaged cool giant exoplanets, largely in preparation for the \emph{Roman Space Telescope} \citep{Lupu2016,Nayak2017,Lacy2019,Damiano2020a,Damiano2020b,Carrion-Gonzalez2020,Carrion-Gonzalez2021,Mukherjee2021}. Subsequent studies have extended reflected light retrievals to directly imaged sub-Neptunes and terrestrial planets \citep{Feng2018,Damiano2021,Damiano2022,Robinson2022,Wang2022}, which would be observable with a future large IR/Optical/UV space-based telescope.

Terrestrial planet reflection spectral retrievals must additionally consider surface reflection. Most previous studies have assumed a uniform-in-wavelength surface albedo \citep{Feng2018,Damiano2022,Robinson2022}. Consequently, the spectral imprint of a wavelength-dependent surface, including the red edge, has seen little investigation. Recently, \citet{Wang2022} found that retrievals using a three-albedo model can outperform a single-albedo model --- demonstrating that reflection spectra can constrain wavelength-dependent surface albedos  \citep[see also][]{Brandt2014}. \citet{Wang2022}'s albedo retrieval technique considered three fixed wavelength bands at visible wavelengths (blue, green, and red). Here, we introduce a generalized albedo retrieval technique designed to identify sharp albedo changes at \emph{a priori} unknown wavelengths analogous to Earth's red edge.

In this study, we demonstrate that wavelength-dependent surface albedos can be retrieved from moderate-quality reflection spectra of Earth-like exoplanets. In particular, the wavelength location of Earth's vegetative red edge can be precisely constrained by a future large IR/Optical/UV space-based telescope. In what follows, we first introduce our modeling and retrieval methodology in Section~\ref{sec:methods}. We demonstrate that our parametric albedo prescription provides an excellent fit to reflection spectra for an Earth-like surface in Section~\ref{sec:results1}. We explore the sensitivity of surface albedo retrievals to data quality and clouds in Section~\ref{sec:results2}, before investigating constraints on other atmospheric and planetary properties in Section~\ref{sec:results3}. Finally, in Section~\ref{sec:discussion}, we summarize our results and discuss their implications.

\section{Reflection Spectra Modeling and Retrieval Framework} \label{sec:methods}

This paper investigates the surface, bulk planetary, and atmospheric properties that can be retrieved from observed reflection spectra of an exo-Earth. We begin in Section~\ref{subsec:models} by presenting a self-consistent 1D model for an Earth-like exoplanet orbiting a Sun-like star and describe the computation of its reflection spectrum. We then outline the generation of synthetic reflection spectral observations and our Bayesian retrieval method for Earth-like exoplanet reflection spectra in Section~\ref{subsec:retrieval_framework}.

\subsection{Reflection Spectra for an Exo-Earth} \label{subsec:models}

\subsubsection{Atmospheric Model} \label{subsubsec:atmos_model}

\begin{figure*}[ht!]
    \centering
    \includegraphics[trim=0cm 0cm 0cm 0.2cm, width=\textwidth]{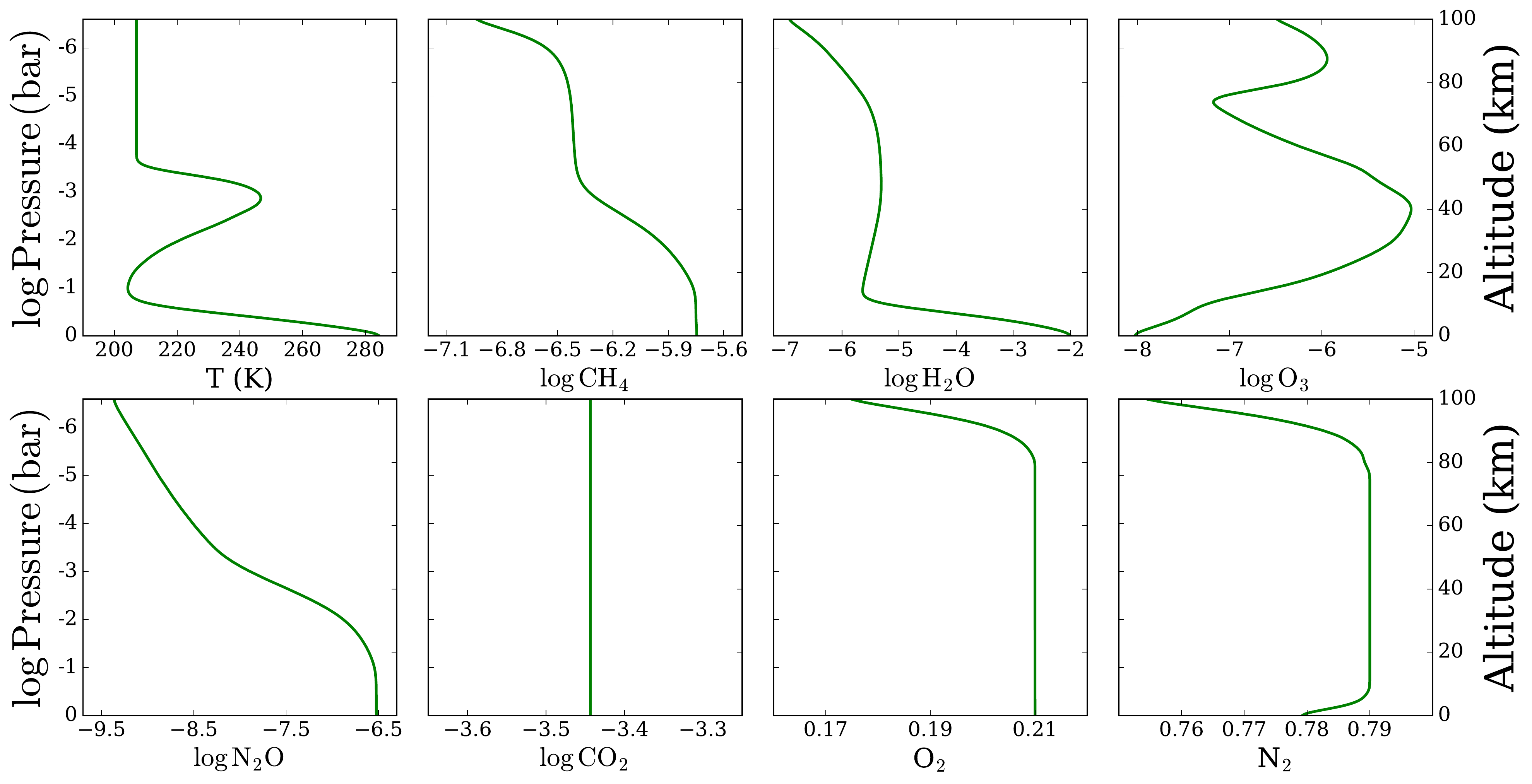}
    \caption{Pressure-temperature profile and volume mixing ratio profiles for a model of the modern Earth. The main spectrally relevant gases from 0.3--2.5\,$\micron$ are shown, alongside the bulk \ce{N2} fraction.}
    \label{fig:atmosmodel}
\end{figure*}

We generate an atmospheric model resembling the modern Earth using Exo-Prime2  \citep[see e.g.,][]{Kaltenegger2010,Madden20,Kasting1986,Pavlov2000,Pavlov02, Segura2005,Segura2007} -- a 1D radiative-convective terrestrial atmosphere code. Exo-Prime2 couples 1D climate and photochemistry models to compute the vertical temperature structure and atmospheric mixing ratio profiles for a planet, assuming an incident stellar spectrum and planetary outgassing rates. Exo-Prime2 also includes feedback from wavelength-dependent surface albedos and clouds. For Earth-like clouds, we use the MODIS 20\,$\micron$ cloud albedo model \citep{King1997,RossowSchiffer1999}, which provides a reasonable average for many clouds of different droplet size. The application of Exo-Prime2 to model Earth-like planets around different host stars and through geological time has been extensively described in the literature \citep[e.g.,][]{Kaltenegger2010,Rugheimer2013,Rugheimer&Kaltenegger2018,Madden20,Lin2022,Kaltenegger2021}. The resulting pressure-temperature (P-T) and mixing ratio profiles computed by Exo-Prime2 are shown in Figure~\ref{fig:atmosmodel}.

\subsubsection{Surface Model} \label{subsubsec:surf_model}

\begin{figure*}[ht!]
    \centering
    \includegraphics[width=0.86\textwidth]{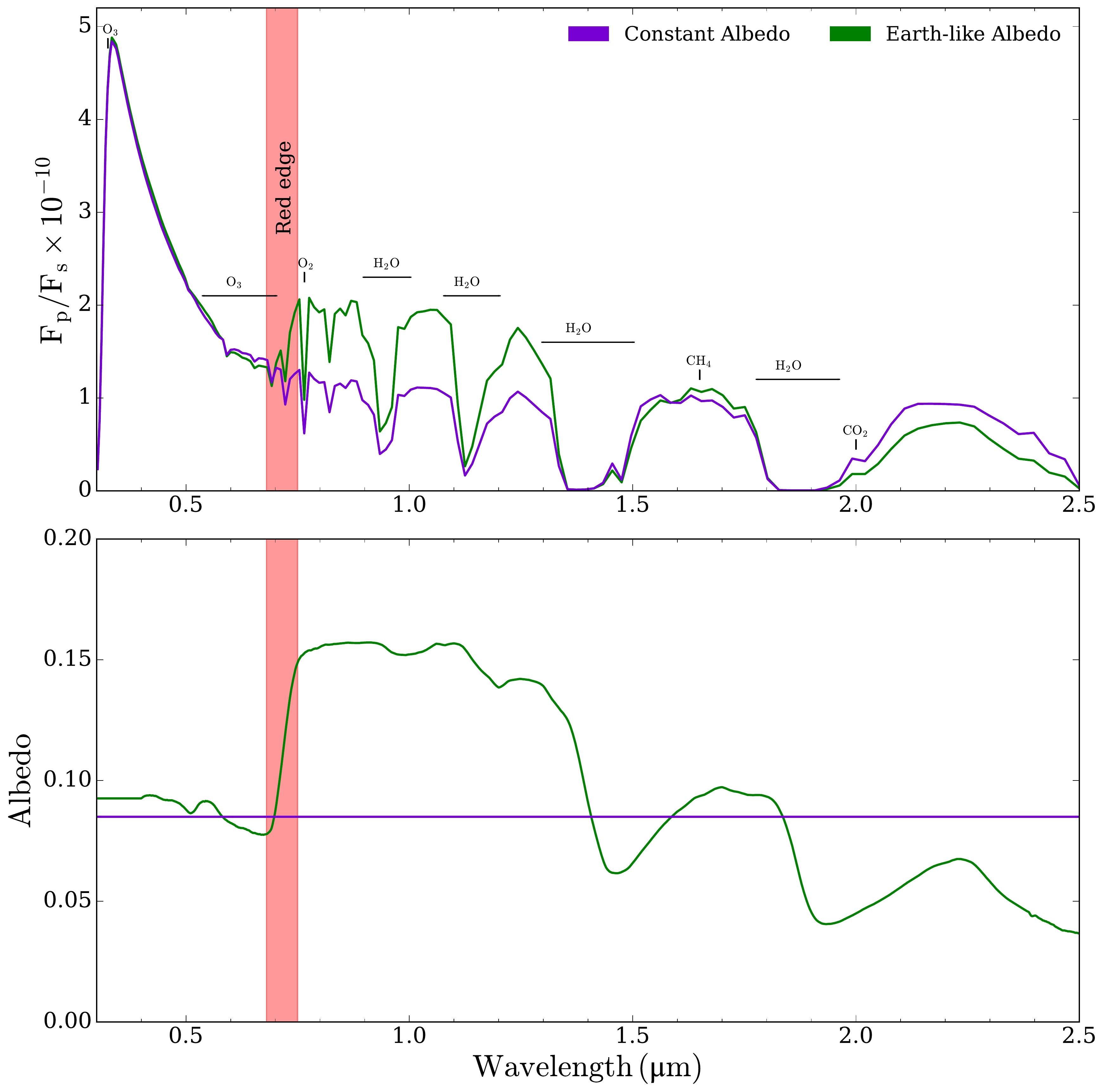}
    \caption{Impact of the red edge on reflection spectra. \textit{Top:} model reflection spectra for a cloud-free Earth-like planet orbiting a Sun-like star. A realistic wavelength-dependent surface albedo (green) produces a sharp increase in the flux ratio around 0.75\,$\micron$ compared to a constant surface albedo (purple). Prominent molecular absorption features are labeled. \textit{Bottom:} the corresponding surface albedos. The location of the vegetative red edge is highlighted in both panels (red shading). For clarity, both models are shown at a spectral resolution of $R=70$.}
    \label{fig:spectra}
\end{figure*}

We model a representative Earth-like surface using wavelength-dependent albedos from the USGS and ASTER spectral libraries \citep{Baldridge2009,Kokaly2017usgs,Clark2007usgs}. We create an average present-day Earth surface albedo from 8 raw albedos of snow, water, coast, sand, trees, grass, basalt, and granite \citep[after][]{Kaltenegger07}. We assume an Earth-like surface consisting of 70\% ocean, 28\% land, and 2\% coast. The land surface consists of 30\% grass, 30\% trees, 9\% granite, 9\% basalt, 15\% snow, and 7\% sand.  We use the surface-fraction weighted albedo (see Figure~\ref{fig:spectra}, bottom panel) in our 1D radiative transfer models.

\subsubsection{Reflection Spectra Computation} \label{subsubsec:spectra_compute}

A distant observer directly imaging an exoplanet measures the wavelength-dependent planet-star flux ratio. At wavelengths where reflected light dominates over thermal emission, the flux ratio can be expressed as

\begin{equation}
    \frac{F_p (\alpha,\lambda)}{F_s (\lambda)} = A_g (\lambda) \, \Phi(\alpha, \lambda)\left( \frac{R_p}{d}\right)^2
\end{equation}
where $A_g$ is the planet's geometric albedo spectrum, $\Phi$ is the phase function, $\alpha$ is the orbital phase, $R_p$ is the planetary radius, and $d$ is the planet-star orbital distance. The geometric albedo is traditionally defined as the ratio of the observed flux from the planet at full phase to that from a perfectly reflecting Lambert disk. The phase function encodes the dilution of the planetary brightness for phase angles without full illumination ($\Phi = 1$ when $\alpha=0$). While the geometric albedo encodes information about an atmosphere's composition, temperature, cloud properties, and surface reflection, the phase function is controlled by both the stellar illumination and atmospheric scattering.

We generate model reflection spectra for an Earth-like planet around a Sun-like star using the open-source radiative transfer code PICASO \citet{Batalha2019}. To compute $F_p / F_s$ from the geometric albedo we provide PICASO with $R_p$ and $d$ (fixed to 1 AU) and assume observations at full phase ($\Phi = 1$) unless otherwise noted. We note that observations will more typically occur at non-zero phase, which would dampen the resultant reflection spectra. However, in this proof of concept study, we choose to focus on full phase to reduce the complexity of the radiative transfer calculations required within the retrieval framework. We set the planetary reference radius such that $r(P = 1\,\rm{bar}) = R_{\Earth}$, the surface at 1\,bar, and the surface gravity to 9.81$\mathrm{ms^{-2}}$. For the stellar spectrum, we used PICASO to interpolate the \citet{Castelli03} grid for a Sun-analog star with $T_{\rm{eff}}$ = 5780\,K, $log \, g$ = 4.437, and [Fe/H] = 0.0122.

For the radiative transfer calculation, we provide PICASO with the P-T profile, mixing ratio profiles, and the wavelength-dependent surface albedo from our Earth-like Exo-Prime2 model. Our reflection spectra computations span the near-UV to near-IR, ranging from 0.3--2.5\,$\micron$. We consider molecular line opacity for \ce{H2O, O2, O3, CH4, CO2, and N2O} \citep[see][for details on the opacity database]{Batalha2019}, alongside Rayleigh scattering from \ce{N2 and O2}. For computational efficiency, we downsampled PICASO's molecular cross sections by 10$\times$ (from R=10,000). We tested different resampling factors and find that $10\times$ downsampling provides a reliable balance between speed and accuracy.

Our PICASO model accounting for Earth-like clouds assumes optical properties consistent with water. Specifically, we use an asymmetry factor of 0.85 and a single scattering albedo of 1.0 \citep[after][]{Feng2018}. We place the cloud base (in log10 bars) at $\log \, p_b=-0.23$, set its vertical extent (also in log10 bars) to $\log \, dp = -0.53$, and the cloud optical depth to $\log \tau = -1.0$. We selected these values for the cloud vertical extent by calibrating the continuum flux from 0.4--1\,$\micron$ of our 1D models to reproduce the reflection spectrum of Earth from \citet{Robinson2011}. We note that assuming a constant cloud albedo increases the reflected flux beyond 1\,$\micron$ compared to \citet{Robinson2011}, but does not significantly influence our analysis.

\subsubsection{Impact of the Red Edge on Reflection Spectra} \label{subsubsec:red_edge_signature}

Figure~\ref{fig:spectra} shows example reflection spectra for a cloud-free exo-Earth, both with and without a wavelength-dependent surface albedo. As expected, the red edge has a pronounced effect on the reflection spectrum. The red edge causes a marked increase in $F_p/F_*$ for wavelengths immediately following 0.75\,$\micron$. As we will see in Section~\ref{sec:results1}, this sudden change is a distinct feature enabling the spectroscopic detectability of the red edge. We also note that the red edge occurs near strong \ce{O3} and \ce{O2} features, which underscores the importance of accounting for a wavelength-dependent surface albedo when searching for atmospheric biosignature gases. With our `ground-truth' model described, we proceed to outline our retrieval framework.

\subsection{Retrieval Framework} \label{subsec:retrieval_framework}

We have developed a Bayesian retrieval wrapper around the PICASO radiative transfer code. We employ this retrieval framework in subsequent sections to demonstrate that information on the red edge can be reliably retrieved from reflection spectra of exo-Earths. Here, we describe the simulated data used in our retrievals and our retrieval configuration.

\subsubsection{Simulated Data \& Noise Model} \label{subsubsec:noise}

Our aim is to investigate the retrievability of the red edge as a function of data quality, rather than for a specific future mission architecture. Consequently, we generated several synthetic datasets, for both the cloud-free and cloudy models described in Section~\ref{subsubsec:spectra_compute}, spanning signal-to-noise ratios of $S/N$ = 5, 10, 15, and 20 (at a reference wavelength of 0.55\,$\micron$) and spectral resolutions of $R$ = 70 and 140. We account for wavelength-dependent noise for the simulated observations using a constant spectral resolution noise model scaling relation as done in \citep{Robinson2016,Feng2018}
 
\begin{equation}
	S/N (\lambda) \propto q(\lambda) \, \mathcal{T}(\lambda) \, A_g(\lambda) \, \Phi (\alpha, \lambda) \, B(\lambda) \, \lambda
\end{equation}
where $q$ is the detector quantum efficiency, $\mathcal{T}$ is the throughput, and $B$ is a blackbody representing the parent star \citep[see][]{Feng2018}. We adopt functions for $q$ and $\mathcal{T}$ from the Python package \emph{coronagraph}\footnote{\href{https://github.com/jlustigy/coronagraph.git}{https://github.com/jlustigy/coronagraph.git}}, which is an open source noise simulator for coronagraph-based observations of directly imaged exoplanets \citep[e.g.,][]{Robinson2016,Lustig2019coronagraph}. For the blackbody, we use $T_{\rm{eff}} = 5780$\,K.

When generating each simulated dataset, we do not randomize the placement of each data point by sampling from a Gaussian distribution. Rather, the data are centered on the (true) planet-to-star flux ratio --- corresponding to the model after binning down to the data resolution --- and assigned error bars according to our noise model at the desired $S/N$. We note that running retrievals on a dataset with Gaussian noise can bias the retrieval results, especially for low spectral resolution and S/N \citep[see][]{Feng2018}. However, running retrievals with Gaussian scatter still allows for spectral features to be recovered \citep[see e.g.,][Appendix A]{Lin2021}. To avoid biasing our retrieval results to a specific random noise draw, we run `scatter-free' retrievals, which produce posterior distributions equivalent to the ensemble average over many individual noise instances. We include an example retrieval with Gaussian scatter in Appendix~\ref{Appendix_A}.

\subsubsection{Retrieval Configuration} \label{subsubsec:retrieval_config}

Bayesian retrieval codes repeatedly call a parameterized radiative transfer forward model to identify the range of bulk planetary, atmospheric, and surface properties consistent with a given dataset. Our retrieval framework employs PICASO \citep{Batalha2019} as the radiative transfer forward model and the MultiNest \citep{Feroz2008,Feroz2009,Feroz2019} wrapper PyMultiNest \citet{Buchner2014} for the sampling algorithm used to explore the parameter space.

We parameterize the atmospheric and surface properties using a simplified prescription that captures the salient features shaping terrestrial exoplanet reflection spectra. We parameterize the P-T profile with an isotherm. We assume \ce{H2O, O2, O3, CH4, CO2, and N2O} are the main spectrally-active gases with sufficient abundances to shape the spectrum for Earth-like planets, with each gas ascribed a single free parameter for the uniform-in-altitude volume mixing ratio. We also assume the primary atmospheric gas is \ce{N2}, with its mixing ratio determined by the condition that mixing ratios must sum to one. We prescribe three further free parameters for the planetary radius and gravity (evaluated at 1\,mbar) and the surface pressure. These choices are similar to those made by other reflected-light retrieval studies \citep[e.g.,][]{Feng2018,Damiano2022,Robinson2022,Wang2022}.

Since our simulated observations incorporate an Earth-like wavelength-dependent surface albedo, we propose a new parametric treatment for wavelength-dependent surface albedos

\begin{equation} \label{eq:3}
    A_s(\lambda) = \begin{cases} 
      A_{s,1} & \lambda < \lambda_1 \\
      A_{s,2} & \lambda_1 \leq \lambda \leq \lambda_2 \\
      A_{s,3} & \lambda > \lambda_2 
   \end{cases}
\end{equation}
where $\mathrm{A_{s,1}}$, $\mathrm{A_{s,2}}$, and $\mathrm{A_{s,3}}$ define the surface albedo in three distinct wavelength regions, $\lambda_1$ marks the transition from $\mathrm{A_{s,1}} \rightarrow \mathrm{A_{s,2}}$, and $\lambda_2$ marks the transition from $\mathrm{A_{s,2}} \rightarrow \mathrm{A_{s,3}}$. This surface albedo prescription thus has five free parameters. To avoid discontinuities at $\lambda_1$ and $\lambda_2$, we compute this function on a wavelength grid at $R$ = 1,000 and convolve it with a Gaussian with a standard deviation of 28 wavelength grid spaces (corresponding to 28\,nm at 1\,$\micron$). Our albedo parameterization thus resembles a smoothed double-step function (similar to the function used by \citealt{Taylor2021} to parameterize the single-scattering albedo of clouds in giant planet nightside emission spectra). We shall demonstrate in subsequent sections that the proposed parameterization is sufficiently flexible to capture both the strong wavelength dependence of the red edge and a possible secondary reflectance edge in the infrared (see Figure~\ref{fig:spectra}).

For retrievals including clouds, we add three further parameters: the cloud base pressure, its vertical pressure extent, and optical depth. Following \citet{Feng2018}, we assume water-like clouds with a fixed asymmetry parameter (0.85) and single scattering albedo (1.0). In total, the most complex retrievals we consider thus have a total of 18 free parameters (summarized in Table~\ref{tab:parameters}). We validated our retrieval framework against simulated data from the \citet{Robinson2011} model (see Appendix \ref{Appendix_B}).

\begin{deluxetable}{llll}
	\tablecaption{Free parameters included in our PICASO retrievals.}
	\label{tab:parameters}
    \tablewidth{0pt}
    \tabletypesize{\scriptsize}
	\tablehead{\colhead{Parameter} & \colhead{Description} & \colhead{Reference Value} & \colhead{Prior Range}}
	\startdata
    $\log \hspace{0.04cm} $\ce{O2} & Oxygen mixing ratio & -0.678 & [-10,0] \\ 
    $\log \hspace{0.04cm} $\ce{O3} & Ozone mixing ratio & -6.25 & [-10,-1] \\ 
    $\log \hspace{0.04cm} $\ce{H2O} & Water vapor mixing ratio  & -2.72  & [-10,-1] \\ 
    $\log \hspace{0.04cm} $\ce{CO2} & Carbon dioxide mixing ratio & -3.44  & [-10,-1] \\ 
    $\log \hspace{0.04cm} $\ce{CH4} & Methane mixing ratio & -5.77  & [-10,-1] \\ 
    $\log \hspace{0.04cm} $\ce{N2O} & Nitrous oxide mixing ratio & -6.55  & [-10,-1] \\
    ${\mathrm{\log \hspace{0.04cm} P_0}}$ & Surface pressure & 0.0  & [-2,2] \\
    $\mathrm{R_p}$ & Planet radius at 1 mbar & 1.007  & [0.5,2.0] \\
    $\mathrm{g}$ & Gravity at 1 mbar & 9.66  & [1.0,25] \\ 
    $\mathrm{T}$ & Temperature & 289 & [100,800] \\
    $\mathrm{\lambda_1}$ & Albedo transition point & 0.72  & [0.3,2.5] \\ 
    $\mathrm{\lambda_2}$ & Albedo transition point & 1.40  & [0.3,2.5] \\ 
    $\mathrm{A_{s,1}}$ & Surface Albedo & 0.09  & [0,1] \\
    $\mathrm{A_{s,2}}$ & Surface Albedo & 0.15  & [0,1] \\ 
    $\mathrm{A_{s,3}}$ & Surface Albedo & 0.06  & [0,1] \\ 
    $\mathrm{\log \hspace{0.04cm} p_b}$ & Cloud-base pressure & -0.23 & [-2,2] \\
    $\mathrm{\log \hspace{0.04cm} dp}$ & Cloud width & -0.53 & [-2,2] \\
    $\mathrm{\log \hspace{0.04cm} \tau}$ & Cloud optical depth & -1.0 & [-2,2] \\
	\enddata
	\tablecomments{The reference values for each parameter correspond to either `ground truth' values from the input  model (e.g., planet radius and cloud properties; see Section~\ref{subsec:models}) or representative average values (e.g., mixing ratios and albedo parameters; see Section~\ref{subsubsec:retrieval_config}). All priors are uniform distributions.}
\end{deluxetable}

Our retrieval analysis covers multiple model and data scenarios. First, in Section~\ref{sec:results1}, we evaluate the retrievability of albedo changes for our cloud-free model, since this model has the strongest spectral red edge. We initially ran four retrievals on the simulated data at $(S/N)_{\rm{ref}} = 5, 10, 15, 20$ and $R = 70$, where we parameterize the planet's wavelength-dependent surface albedo with Equation~\ref{eq:3}. Then, we ran a similar set of retrievals with a constant-in-wavelength surface albedo. Doing so enables us to perform Bayesian model comparisons between the wavelength-dependent surface and the constant-in-wavelength surface models \citep[e.g.,][]{Benneke2013,Trotta2017}. We also ran a retrieval with $R = 140$ at $(S/N)_{\rm{ref}} = 10$ to investigate the impact of retrieving data at a higher spectral resolution. Lastly, we ran a retrieval at $R = 70$ and $(S/N)_{\rm{ref}} = 10$ for a more realistic dataset including a cloud deck to investigate how clouds impact the retrieval results. All our MultiNest retrievals use 2,000 live points, which typically involve the computation of $10^6$ models.

We summarize the prior range for each retrieval free parameter in Table~\ref{tab:parameters}. We generally allow generous prior ranges, encompassing a wide range of physically plausible values, with all priors being uniform distributions. As in \citet{Feng2018}, we allow for oxygen-rich atmospheres by extending its prior range to 100\% (but rejecting any parameter combinations where the sum of the non-\ce{N2} mixing ratios exceed unity). Our prior range for the planet radius and gravity terminates at 2.0\,$R_{\Earth}$ and 25.0\,$\mathrm{m s^{-2}}$, respectively, since this study focuses on Earth-like planets. We note that PICASO requires that the pressure corresponding to $R_p$ must be less than the highest pressure in the atmospheric pressure grid (i.e. the surface pressure). We circumvent this issue by defining the planet radius and gravity parameters at 1\,mbar, such that the surface pressure prior range ($10^{-2}$--$10^{2}$\,bar) is always deeper than 1\,mbar.

We also include `reference values' for each parameter in Table~\ref{tab:parameters} --- corresponding closest to the original input Exo-Prime2 model (see Section~\ref{subsec:models}) --- for comparison to the retrieval results. Since the input gas mixing ratios depend on height, while the retrievals assume uniform mixing ratios, we set the reference values as the average of the true mixing ratio profile from the surface to 25\,km altitude. Our reference temperature is the planet's surface temperature. For the albedo parameters, our reference values are determined by averaging the surface albedo over 0.3--0.72\,$\micron$, 0.72--1.4\,$\micron$, and 1.4--2.5\,$\micron$. The reference planet radius and gravity correspond to the true values from the Exo-Prime2 model (scaled to 1\,mbar). Similarly, the surface pressure and cloud parameter reference values correspond exactly to the original model inputs (see Section~\ref{subsubsec:spectra_compute}).

We now turn to present the results of our retrievals including wavelength-dependent surface albedos.

\newpage

\section{The Necessity for Wavelength- dependent Surface Albedos in Reflected-light Retrievals} 
\label{sec:results1}

Here, we demonstrate that reflection spectra of terrestrial exoplanets contain recoverable information on wavelength-dependent surfaces. We show that not only can data commensurate with future direct imaging missions constrain wavelength-dependent surface albedos, but that assuming a constant surface albedo may result in biased atmospheric inferences.

\subsection{Can a Uniform Albedo Fit the Earth's Red Edge?} \label{subsec:spectral_fit_quality}

We first assess whether a retrieval model assuming a constant-in-wavelength surface albedo can adequately fit the reflection spectrum of an exoplanet with a realistic Earth-like surface. We have already seen in Section~\ref{subsubsec:red_edge_signature} and Figure~\ref{fig:spectra} that the Earth's red edge induces a sharp change in $F_p/F_*$ around 750\,nm, so here we quantify whether such a spectral signature is detectable and its impact on atmospheric retrievals.

\begin{figure*}[ht!]
    \centering
    \includegraphics[width=\textwidth]{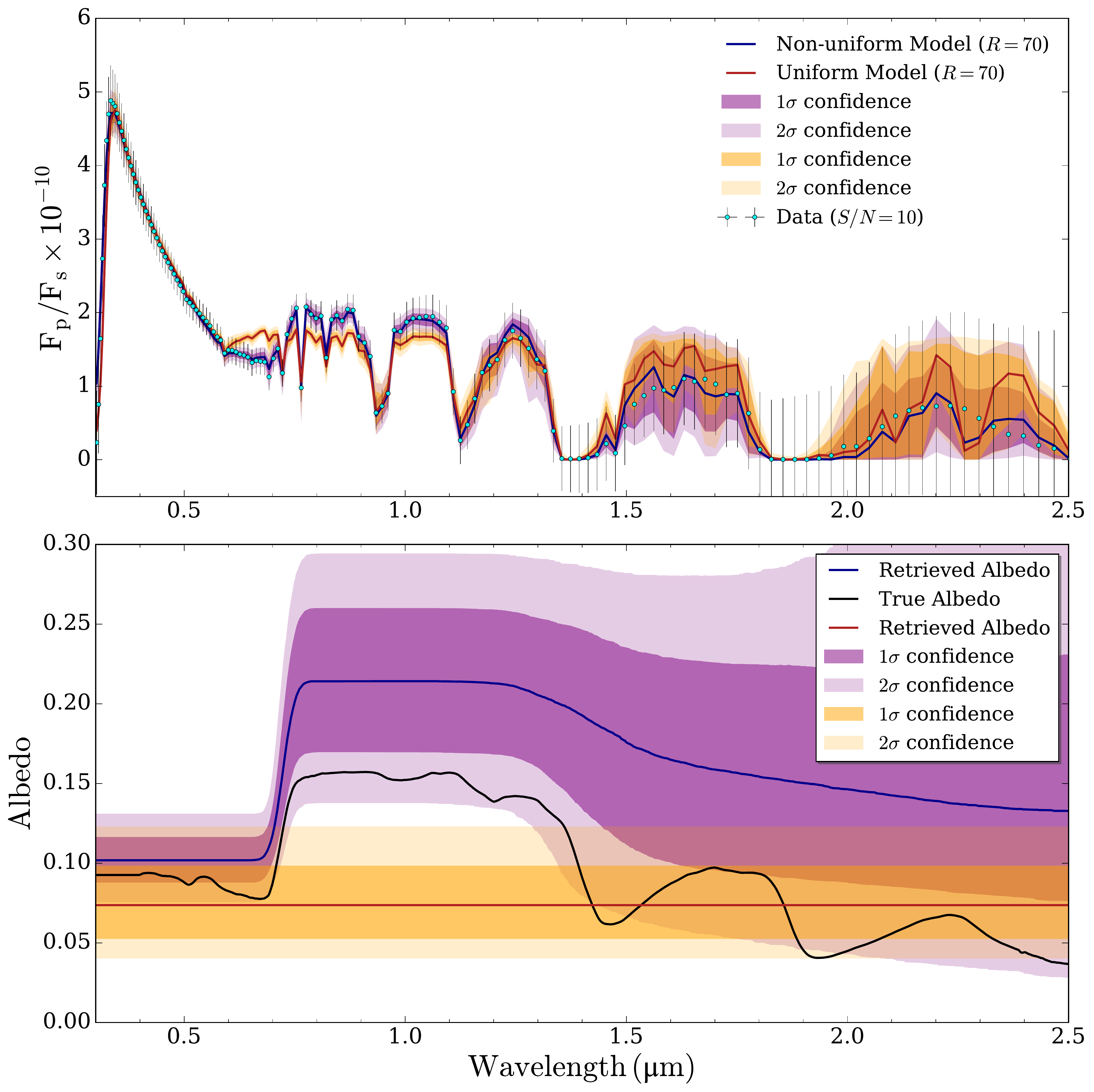}
    \caption{\textit{Top:} reflection spectra retrieval results assuming a wavelength-dependent surface albedo (purple contours) and a uniform surface albedo (orange contours). The simulated data ($R = 70$ and $(S/N)_{\rm{ref}} = 10$) corresponds to an Earth-like exoplanet orbiting a Sun-like star with an Earth-like wavelength-dependent surface. \textit{Bottom:} the retrieved surface albedo profiles corresponding to the two models in the top panel. The true Earth-like surface albedo used to generate the simulated data is overlaid for comparison (black curve).}
    \label{fig:retrievedspectra}
\end{figure*}

In Figure~\ref{fig:retrievedspectra}, we demonstrate that a uniform albedo model often struggles to capture the spectral morphology of an Earth-like exoplanet. Our `ground truth' model is the cloud-free scenario described in Section~\ref{subsec:models}, which produces the strongest red edge, while the simulated data (here, $(S/N)_{\rm{ref}} = 10$) and retrieval configuration are detailed in Section~\ref{subsec:retrieval_framework}. We see that the retrieved spectrum for the uniform albedo model begins to deviate from the simulated data for optical wavelengths longer than 0.65\,$\micron$. In the optical and near-IR, where the S/N is highest, the uniform albedo model is often discrepant with the data to 2$\sigma$. The root cause of this model-data mismatch is that the uniform albedo model has a roughly constant continuum $Fp/F*$ from 0.6--0.9\,$\micron$ (outside \ce{O2} absorption features), which cannot reproduce the sharp spectral continuum change associated with the vegetation red edge near 0.72\,$\micron$.

In contrast, our wavelength-dependent surface albedo parameterization well matches the reflection spectrum of our Earth-like exoplanet. Equation~\ref{eq:3} allows our retrieval code to reproduce both the large increase in $F_p/F_*$ at 0.72\,$\micron$ caused by the vegetation red edge and the general morphology of the spectrum in the visible, and near-IR. Statistically, our retrieval including a wavelength-dependent albedo is favored over the uniform model with a Bayes factor of $\ln \mathcal{B} = 30.4$ (equivalent to 8.1\,$\sigma$ using the relations in \citealt{Benneke2012}), which would be considered a conclusive detection on the Jeffrey's scale of Bayesian model comparison \citep[e.g.][]{Trotta2017}.

We illustrate why the uniform albedo model struggles to fit our data by comparing the retrieved surface albedos in the bottom panel of Figure~\ref{fig:retrievedspectra}. While the uniform albedo model correctly captures the surface albedo in the near-IR wavelengths beyond 1.4\,$\micron$, it significantly underestimates the true surface albedo from 0.75--1.35\,$\micron$. However, our proposed parameterization demonstrates that one can retrieve wavelength-dependent surface properties --- in particular the location of the vegetation red edge --- at even a moderate signal-to-noise ratio ($(S/N)_{\rm{ref}} = 10$). The retrieved surface albedo profile also correctly infers a decrease in the albedo for near-IR wavelengths beyond 1.3\,$\micron$. While our wavelength-dependent surface retrieval somewhat overestimates the magnitude of the albedo (likely due to other complexities not captured in the model, such as variable atmospheric abundances with height), it correctly captures the general shape of the wavelength-dependent albedo profile and lies within 2$\sigma$ of the true Earth-like surface albedo.

\begin{figure}[t!]
    \centering
    \includegraphics[trim=0.0cm 13cm 0.0cm 0.0 cm,width=\columnwidth]{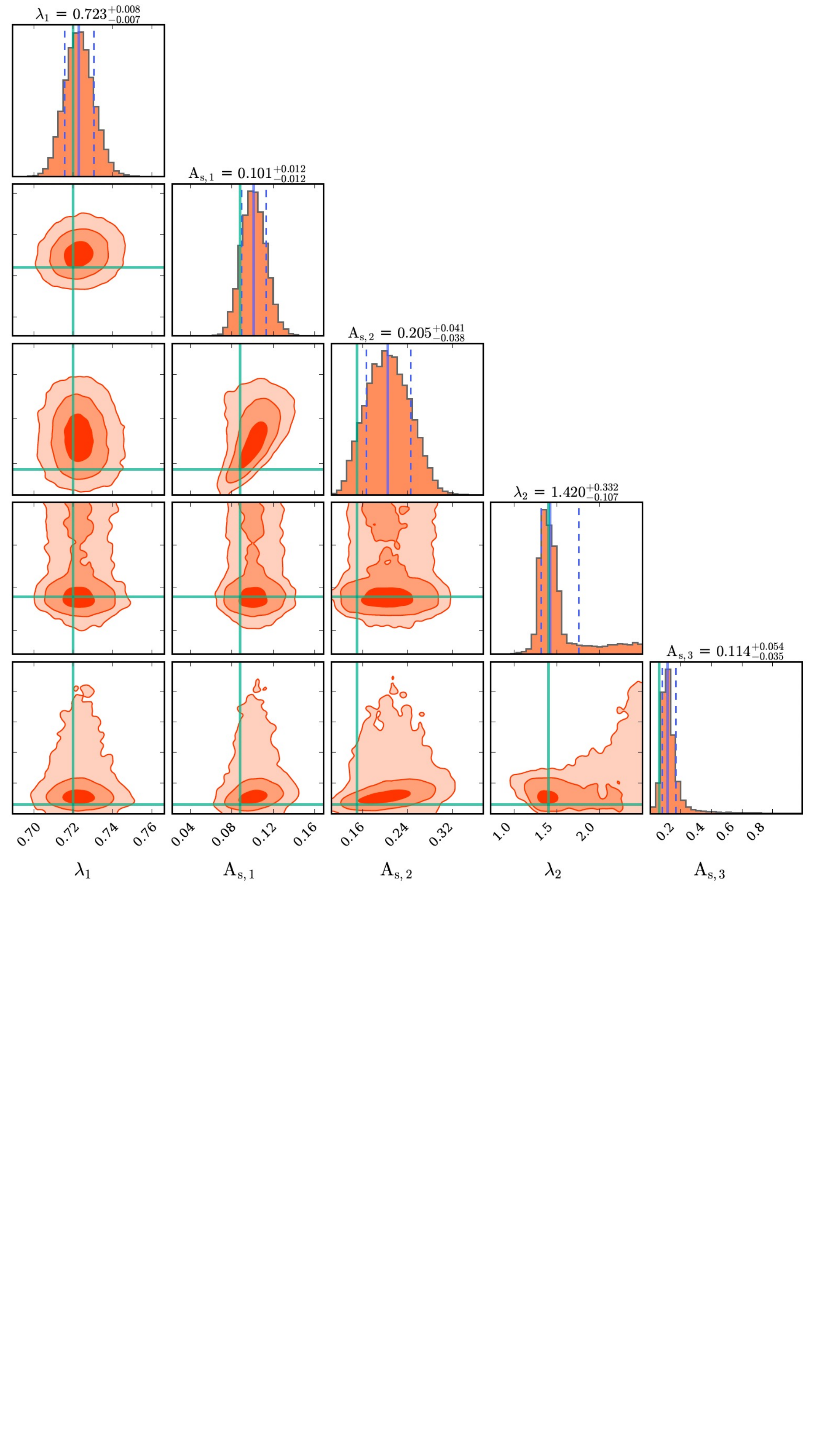}
    \caption{Posterior probability distribution for the retrieved surface albedo parameters in Equation~\ref{eq:3} (for simulated data at $R = 70$ and $(S/N)_{\rm{ref}} = 10$). The green lines mark the ground truth references values (see Table~\ref{tab:parameters}). The orange contours show the 1$\sigma$, 2$\sigma$, and 3$\sigma$ confidence regions for each retrieved parameter. The blue lines in the histograms show the median (solid line) and marginalized 1$\sigma$ confidence region (dashed lines) for each parameter.}
    \label{fig:surface_params}
\end{figure}

\begin{figure}
    \centering
    \includegraphics[width=\columnwidth]{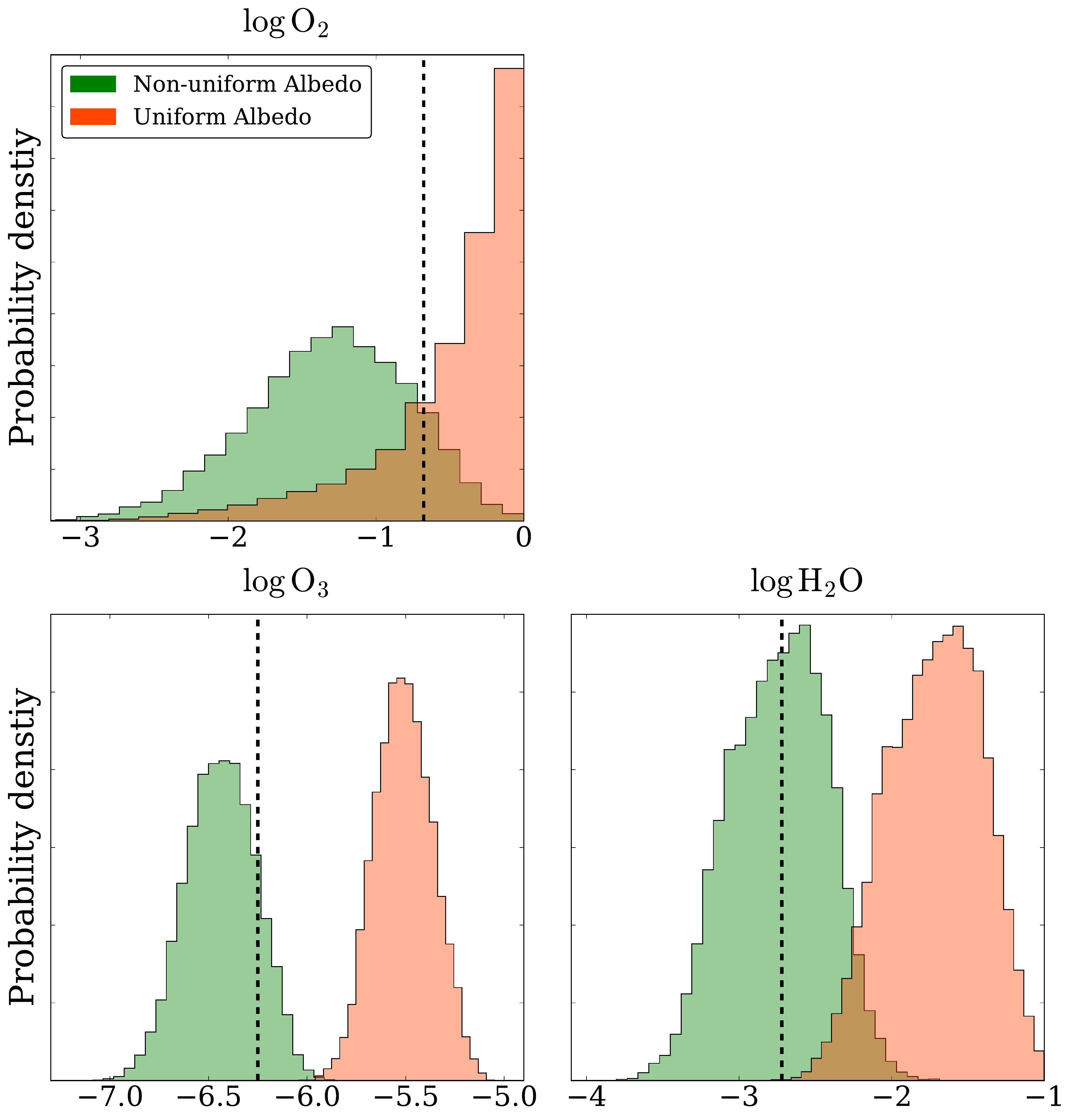}
    \caption{Impact on the retrieved molecular abundances of \ce{O2, O3, and H2O} from assuming a uniform surface albedo. Accounting for a wavelength-dependent surface (green histograms) results in good agreement with the ground truth reference values (black lines, see Table~\ref{tab:parameters}), while a retrieval assuming a uniform albedo (orange) can bias these abundances by an order of magnitude.}
    \label{fig:atmospheric_biases}
\end{figure}

We further show the posteriors of our retrieved albedo parameters in Figure~\ref{fig:surface_params}. All five parameters are well-constrained by the data, demonstrating that reflection spectra are highly sensitive to the wavelength-dependence of the surface albedo. In particular, the parameter encoding the wavelength location of the red edge, $\lambda_1$, is retrieved to a remarkable precision of 8\,nm. Similarly, the retrieval identifies a secondary albedo change near 1.4\,$\micron$, encoded by $\lambda_2$, with a precision of $\approx$ 200\,nm (though the long posterior tail to higher values indicates this is harder to constrain than $\lambda_1$). The three albedo parameters are slightly overestimated, as noted above, but are consistent within 2\,$\sigma$ of the reference values (see Table~\ref{tab:parameters}). Overall, Figure~\ref{fig:surface_params} shows that Equation~\ref{eq:3} offers a parametrization capable of capturing the key wavelength-dependent features of a realistic Earth-like surface.

\newpage

\subsection{Biases from Assuming a Uniform Albedo} \label{subsec:uniform_albedo_biases}

The assumption of a uniform surface albedo can bias inferred properties of an exoplanet. Since a retrieval code employs every available means to minimize model-data residuals, it can attempt to compensate for the non-inclusion of a wavelength-dependent surface by modifying the retrieved abundances of chemical species in the atmosphere (since their cross sections are also wavelength dependent). In Figure~\ref{fig:atmospheric_biases}, we demonstrate that one consequence from assuming a uniform surface albedo is biased abundance inferences for several key molecules in our model. Specifically, we find that the volume mixing ratios of \ce{O3 and H2O} are over-estimated by an order of magnitude and the bulk atmospheric gas would be identified as \ce{O2} rather than \ce{N2}. This finding underscores an important point: accurate abundance inferences for atmospheric gases can depend on the inclusion of a wavelength-dependent surface albedo in reflected light retrieval frameworks. Since Figure~\ref{fig:atmospheric_biases} corresponds to the moderate case of $R = 70$ and $(S/N)_{\rm{ref}} = 10$, wavelength-dependent surface spectral properties will be an important consideration for future direct imaging missions for exoplanets, especially for missions focused on Earth-like exoplanets.

\section{Factors Influencing Retrievals of Wavelength-dependent \\ Surface Albedos} \label{sec:results2}

Having established the need for a wavelength-dependent surface albedo model in reflected-light retrievals, we next explore how these results depend on data quality ($S/N$ in Section~\ref{subsec:S_N_sensitivity} and $R$ in Section~\ref{subsec:R_sensitivity}) and the inclusion of clouds (Section~\ref{subsec:cloud_sensitivity}).

\subsection{Sensitivity to S/N}
\label{subsec:S_N_sensitivity}

\begin{figure*}[ht!]
    \centering
    \includegraphics[trim = 0.0cm -1.0cm 0.0cm -1.0cm, width=\textwidth]{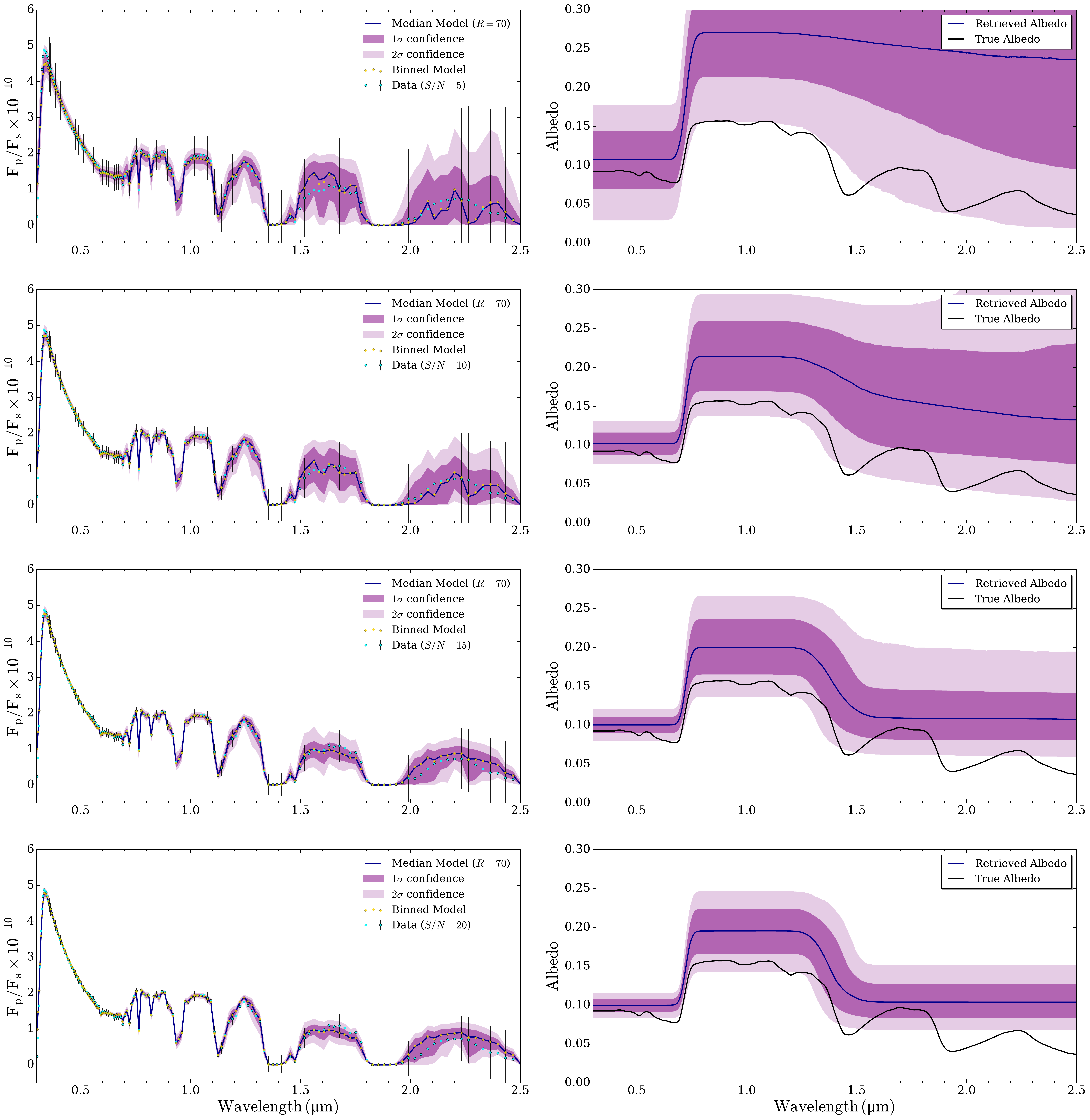}
    \caption{Retrieved reflection spectra and surface albedo profiles as a function of signal-to-noise ratio. \textit{Left panels}: comparison of the median retrieved spectrum (blue line), and its corresponding 1$\sigma$ and 2$\sigma$ confidence regions (purple contours), with simulated data for an Earth-like exoplanet (cyan errors) at $(S/N)_{\rm{ref}}$ = 5, 10, 15, and 20. The median model binned to the resolution of the data (gold diamonds) well fit the data. \textit{Right panels}: corresponding retrieved surface albedo profiles (blue line and purple contours) inferred from each dataset compared to the true Earth-like surface model (black line).}
    \label{fig:retrievedspectra_SNgrid}
\end{figure*}

Figure~\ref{fig:retrievedspectra_SNgrid} shows our retrieved spectra and surface albedo profiles for $(S/N)_{\rm{ref}}$ = 5, 10, 15, and 20. We see that even at $(S/N)_{\rm{ref}} = 5$, the retrieval correctly identifies a sharp rise in the surface albedo near 0.7\,$\micron$ --- consistent with the wavelength of the vegetation red edge on modern Earth. The uncertainty in the retrieved wavelength of this feature is remarkably small ($\lambda_1$ determined to $\approx$ 15\,nm). This suggests that sudden changes in surface albedo are an effect of first order importance even for low signal-to-noise observations. With a doubling to $(S/N)_{\rm{ref}} = 10$, we see further improvements in the retrieved surface albedo profile: (i) the uncertainty in the location of the sharp rise in surface albedo is halved ($\lambda_1$ determined to $\approx$ 8\,nm); (ii) a hint emerges of a secondary albedo change near 1.4\,$\micron$ ($\lambda_2$ determined to $\approx$ 200\,nm); and (iii) the true surface albedo profile is correctly captured throughout most of the wavelength range to within 2$\sigma$. For $(S/N)_{\rm{ref}} = 15$, the retrieval becomes more confident about the existence of a secondary albedo edge ($\lambda_2$ determined to $\approx$ 70\,nm). Finally, at $(S/N)_{\rm{ref}} = 20$ the retrieved model attains even better overall agreement with the true albedo model. We also find that the tendency to overestimate the retrieved albedo (see Section~\ref{subsec:spectral_fit_quality}) becomes less prevalent for higher S/N.

Table~\ref{tab:significances} quantifies the preference for our wavelength-dependent albedo model (Equation~\ref{eq:3}) over a uniform albedo model. For $(S/N)_{\rm{ref}} = 5$, our Bayesian model comparison finds moderate evidence for a non-uniform surface albedo (2.7$\sigma$). A slight increase to $(S/N)_{\rm{ref}} = 10$ suffices to conclusively detect at least one discontinuity (8.1$\sigma$). Further increases in $S/N$ can help to detect a non-uniform surface albedo also for cloudy atmospheres, where the effect is smaller because clouds block part of the light from the underlying surface from view (see Section~\ref{subsec:cloud_sensitivity}). Since our retrievals thus far have only considered data at $R = 70$, we next explore variable spectral resolution for a fixed signal-to-noise ratio.

\subsection{The Role of Spectral Resolution}
\label{subsec:R_sensitivity}

\begin{figure}[t!]
    \centering
    \includegraphics[trim = 1.0cm -0.5cm 1.0cm -0.5cm, width=\columnwidth]{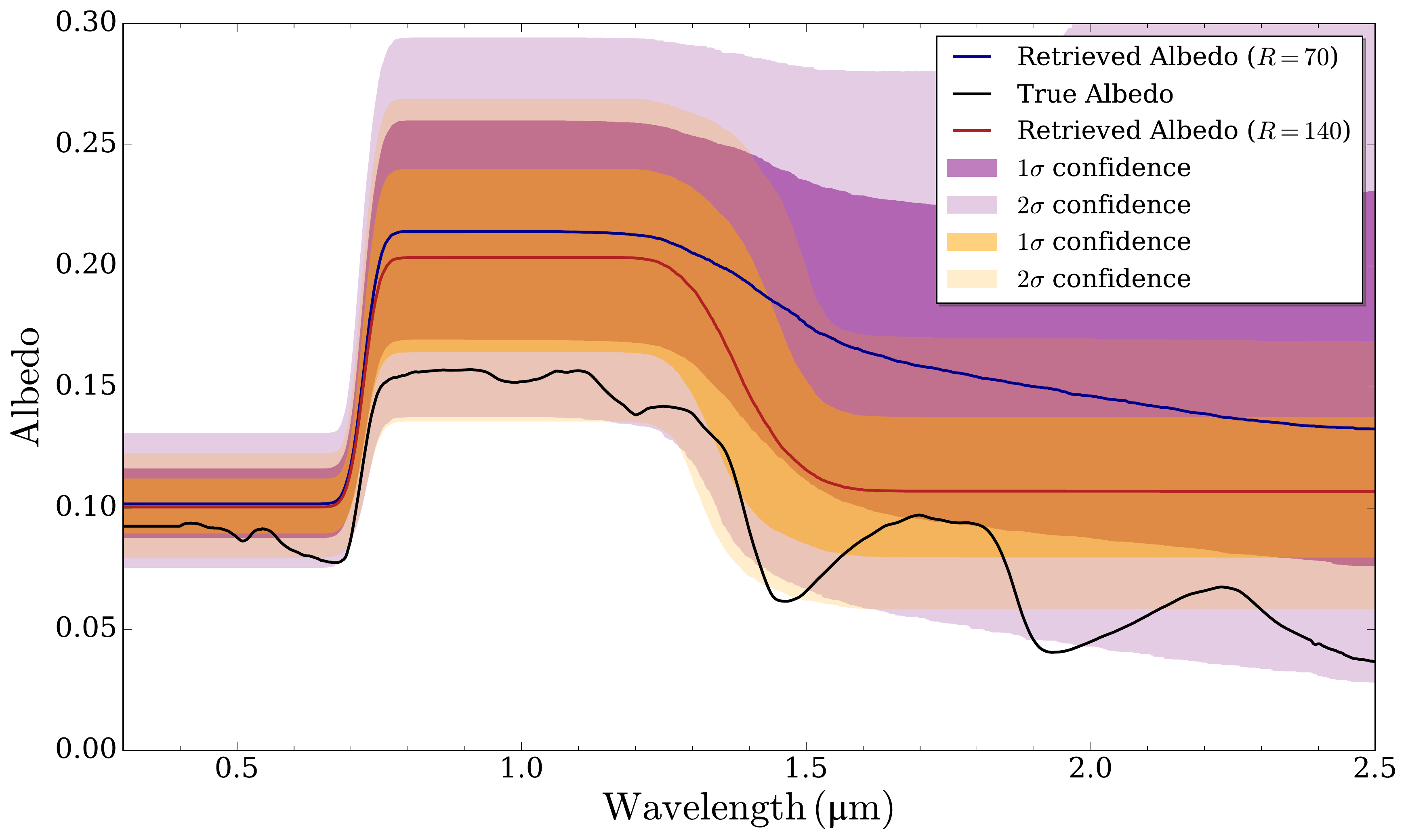}
    \caption{Impact of spectral resolution on the retrieved surface albedos of modern Earth seen as an exoplanet. The retrieved albedos for two distinct datasets are overlaid: (i) $R = 70$ and $(S/N)_{\rm{ref}} = 10$ (purple contours); and (ii) $R = 140$ and $(S/N)_{\rm{ref}} = 10$ (orange contours). The true model Earth surface used to generate both simulated datasets is included for comparison (black line).}
    \label{fig:retrievedspectra_resolution}
\end{figure}

Figure~\ref{fig:retrievedspectra_resolution} shows how the retrieved surface albedo profile changes with spectral resolution. Specifically, we illustrate the expected improvement from doubling the spectral resolution from $R = 70$ to $R = 140$ (hence doubling the number of datapoints from 0.3--2.5\,$\micron$). We see that the retrieved albedo from the higher resolution data is in better agreement with the true albedo, especially at longer wavelengths where the errors bars are largest. Further, the uncertainty on the retrieved albedo parameters decrease ($\mathrm{A_{s,1}}$ and $\mathrm{A_{s,2}}$ improve by $\approx$ 25\%; $\mathrm{A_{s,3}}$ improves by $\approx$ 50\%). We further note that the at $R = 140$ the retrieval of the second, smaller edge around 1.4\,$\micron$ ($\lambda_2$) improves (by $\approx$ 68\%), resulting in a retrieved albedo shape more consistent with the true Earth-like surface.

\subsection{The Impact of Clouds}
\label{subsec:cloud_sensitivity}

\begin{deluxetable}{lllll}[t!]
	\tablecaption{Predicted detection significances for an Earth-like wavelength-dependent surface albedo as a function of S/N.}
	\label{tab:significances}
	\tablewidth{0pt}
	\tablehead{\colhead{Model} & \colhead{$(S/N)_{\rm{ref}}=5$} & \colhead{$(S/N)_{\rm{ref}}=10$} & \colhead{$(S/N)_{\rm{ref}}=15$} & \colhead{$(S/N)_{\rm{ref}}=20$}}
	\startdata
	Clear & 2.7$\sigma$ & 8.1$\sigma$ & 12.7$\sigma$ & 17.2$\sigma$ \\
    Cloudy & --- & 2.9$\sigma$ & 5.6$\sigma$ & --- \\
	\enddata
	\tablecomments{$(S/N)_{\rm{ref}}$ is defined at 0.55\,$\micron$ (see Section~\ref{subsubsec:noise}). All data is at $R = 70$ from 0.3--2.5\,$\micron$. Cloudy spectra retrievals at $(S/N)_{\rm{ref}}=5$ and $(S/N)_{\rm{ref}} = 20$ were not computed, so `---' is used for their detection significances.}
\end{deluxetable}

\begin{figure}
    \centering
    \includegraphics[trim = 1.0cm -0.5cm 1.0cm -0.5cm, width=\columnwidth]{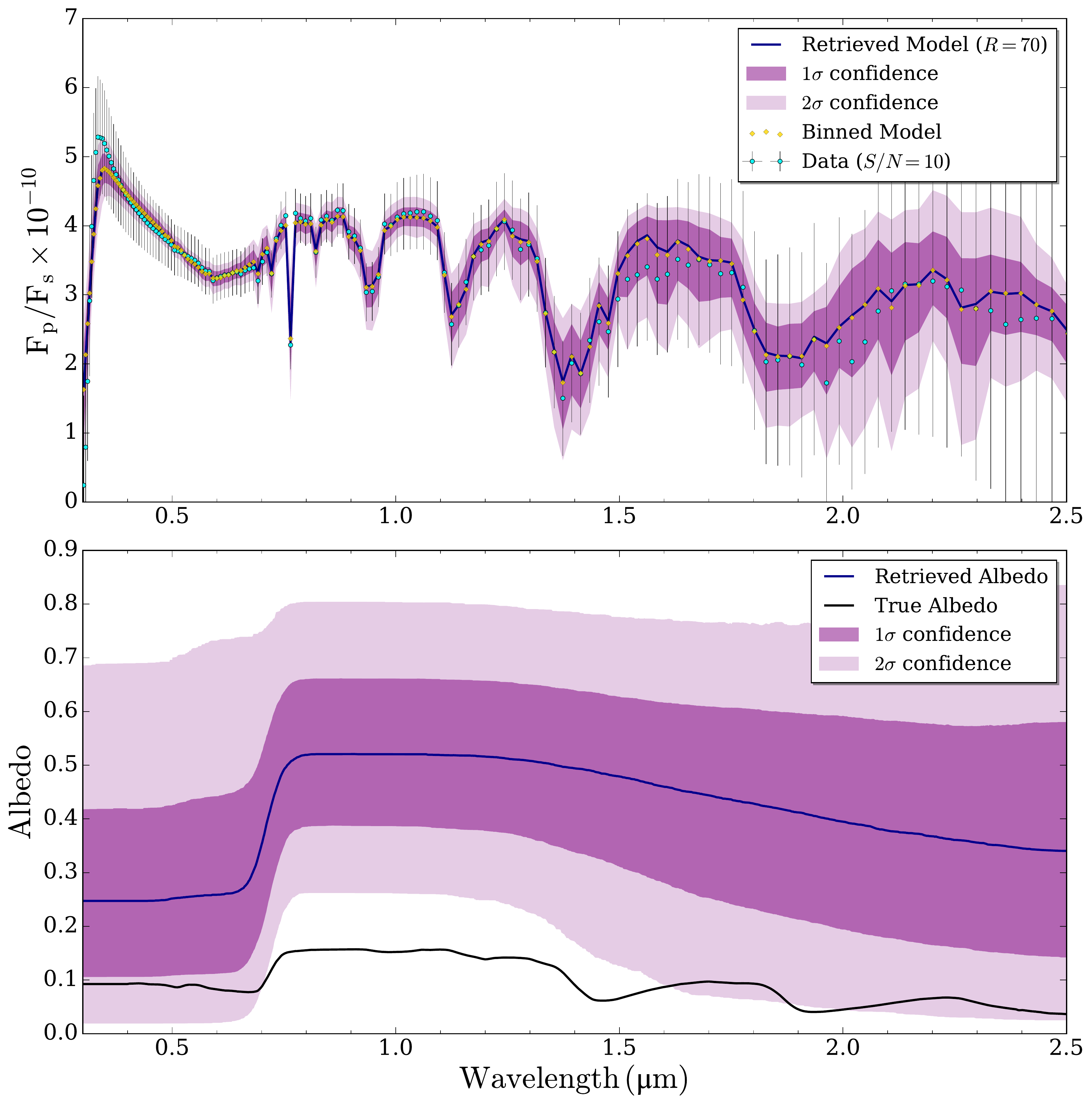}
    \caption{\textit{Top:} retrieved reflection spectrum of a cloudy Earth-like exoplanet with a wavelength-dependent surface albedo (for simulated data at $R = 70$ and $(S/N)_{\rm{ref}} = 10$). The median retrieved spectrum binned to the resolution of the data (gold diamonds) generally agrees with the simulated observations (cyan errors) within the retrieval confidence intervals (purple contours). \textit{Bottom:} corresponding retrieved surface albedo profile (purple contours) compared to the true surface albedo (black line).}
    \label{fig:retrievedcloudyspectra}
\end{figure}

Figure~\ref{fig:retrievedcloudyspectra} shows how the inclusion of clouds affects the retrieved spectrum and surface albedo profile. Our cloud properties for this demonstration were chosen to resemble the continuum flux from the model in \citet{Robinson2011} (see Section~\ref{subsubsec:atmos_model}). We see that while the data is well fit, the retrieved albedo profile is generally overestimated when clouds are included (see also \citealt{Wang2022} for a discussion on cloud-surface degeneracies). Nevertheless, the wavelength of the red edge is still reliably retrieved and well constrained even in the presence of clouds ($\lambda_1$ determined to 17\,nm). The retrieved albedo also shows a slight decrease at longer wavelengths, but the secondary albedo change near 1.4\,$\micron$ is not well constrained. Overall, the presence of a cloud deck can lead one to infer an artificially brighter surface outside the 1$\sigma$ uncertainty region of the retrieved albedo profile.

Clouds increase the minimum signal-to-noise ratio required to detect a wavelength-dependent surface feature. For example, Table~\ref{tab:significances} demonstrates that the detection significance for a non-uniform surface albedo at $(S/N)_{\rm{ref}} = 10$ drops from 8.1$\sigma$ (cloud-free) to 2.9$\sigma$ (including clouds). Such lower significances arise from cloud-surface degeneracies broadening albedo uncertainties (see Figure~\ref{fig:retrievedcloudyspectra}). However, we still find a detection of a wavelength-dependent surface albedo for $(S/N)_{\rm{ref}} = 15$ (5.6$\sigma$). These results show that, while clouds can complicate the inference of wavelength-dependent surface features, it is still possible to identify non-uniform surface albedos for Earth-like cloud coverage.

\newpage

\section{Full Retrieval Results for Planetary and Atmospheric Properties}
\label{sec:results3}

\begin{figure*}[ht!]
    \centering
    \includegraphics[width=0.98\textwidth]{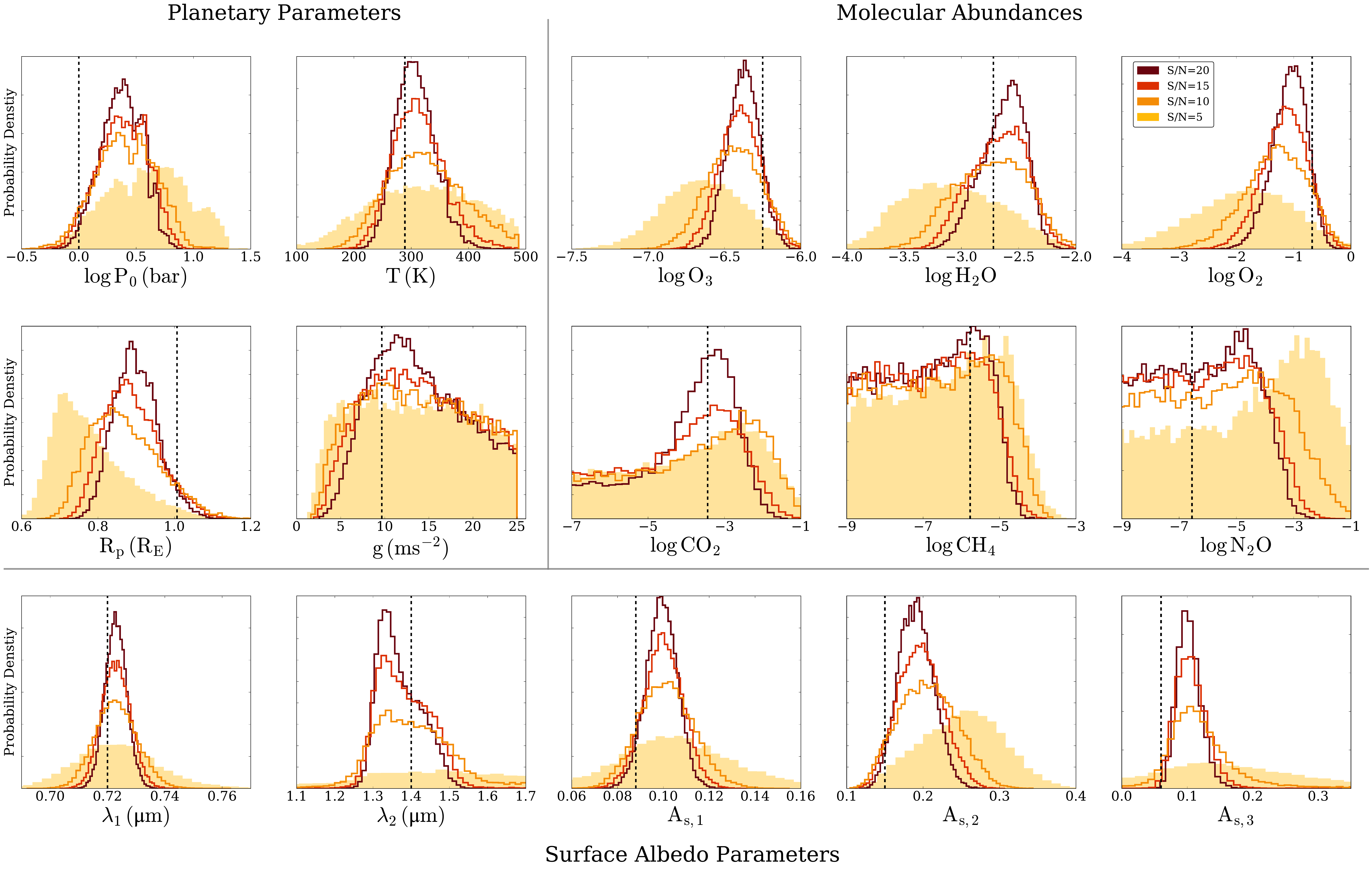}
    \caption{Retrieval results for bulk planetary, atmospheric, and surface parameters for a cloud-free Earth-like planet. Four different retrieval results are shown as a function of S/N (shading and histogram outlines). The reference values for the ground truth model are annotated (black dashed lines).}
    \label{fig:retrievedhistograms}
\end{figure*}

For completeness, here we show our full retrieval results for other planetary and atmospheric properties. Table~\ref{tab:retrievalresults} summarizes the retrieved values of all 15 free parameters for our cloud-free scenario, along with their 1$\sigma$ uncertainties, as a function of S/N. We also show the posterior distributions for each parameter in Figure~\ref{fig:retrievedhistograms}.

\subsection{Planetary Properties}

Figure~\ref{fig:retrievedhistograms} (top left) shows that the reliable inference of bulk planetary properties requires moderate signal-to-noise ratios for cloud-free models. At $(S/N)_{\rm{ref}} = 5$, the 1\,mbar radius is underestimated and the surface pressure slightly overestimated. The temperature is correctly retrieved, albeit with broad uncertainty ($\approx$ 100\,K). We find that $(S/N)_{\rm{ref}} = 10$ is the minimum to reliably retrieve these parameters. In particular, for the temperature a well-defined peak appears around 300\,K and the uncertainty shrinks to $\approx$ 70\,K. The surface pressure and planetary radius are less biased for $(S/N)_{\rm{ref}} \geq 10$, with the reference values correctly retrieved within 2$\sigma$. For the 1 mbar gravity, we find only a lower limit for all our signal-to-noise ratios.
 
\subsection{Molecular Abundances}

Figure~\ref{fig:retrievedhistograms} (top right) shows that the abundances of gases with strong absorption features in the optical and near-infrared (\ce{O3, O2}, and \ce{H2O}) are generally well constrained. The \ce{O3} abundance can always be constrained better than a factor of 2 (0.3\,dex), even for $(S/N)_{\rm{ref}} = 5$. The ease of constraining \ce{O3} is driven by its strong absorption at optical and near-UV wavelengths. The \ce{O2} posterior is the broadest due to the smaller number of data points spanning its narrow absorption features. Nevertheless, \ce{O2} can be constrained to 0.3\,dex for $(S/N)_{\rm{ref}} = 20$. We note that the \ce{O3}, \ce{O2}, and \ce{H2O} abundances are slightly underestimated for $(S/N)_{\rm{ref}} = 5$, but are reliably retrieved for $(S/N)_{\rm{ref}} \geq 10$.

Our retrievals are unable to detect gases with only weak absorption features in the modeled wavelength range. We can place an upper limit on the \ce{CH4} abundance for $(S/N)_{\rm{ref}} = 5$, but \ce{CO2} and \ce{N2O} require $(S/N)_{\rm{ref}} = 15$ for upper limits. We find a tentative hint of \ce{CO2} at $(S/N)_{\rm{ref}} = 20$ centered on the reference value, but the posterior tail to lower abundances indicates a non-detection of \ce{CO2} absorption. Constraints on gases such as \ce{CO2} and \ce{CH4} for a modern Earth-like atmosphere at low resolution and low signal-to-noise would benefit from observations of thermal emission in the mid-infrared \citep[e.g.][]{DesMarais2002,Kaltenegger07,Konrad2021}.

\begin{deluxetable}{llrrrr}
    \tablecaption{Retrieval results for a clear atmosphere at R=70 with reference values and 1$\sigma$ uncertainties.}
    \label{tab:retrievalresults}
	\tablewidth{0pt}
	\tabletypesize{\scriptsize}
	\tablehead{\colhead{Parameter} & \colhead{Reference} & \colhead{S/N=5} & \colhead{S/N=10} & \colhead{S/N=15} & \colhead{S/N=20}}
	\startdata
    $\log \hspace{0.04cm} $\ce{O2} & -0.678 & 
    $-1.87 \substack{+0.85 \\ -1.02}$ & 
    $-1.29 \substack{+0.52 \\ -0.57}$ & 
    $-1.14 \substack{+0.37 \\ -0.42}$ & 
    $-1.06 \substack{+0.28 \\ -0.32}$\\
    $\log \hspace{0.04cm} $\ce{O3} & -6.25 & 
    $-6.64 \substack{+0.27 \\ -0.27}$ & 
    $-6.44 \substack{+0.18 \\ -0.18}$ & 
    $-6.40 \substack{+0.14 \\ -0.14}$ & 
    $-6.37 \substack{+0.10 \\ -0.11}$\\
    $\log \hspace{0.04cm} $\ce{H2O} & -2.72 & 
    $-3.09 \substack{+0.48 \\ -0.43}$ & 
    $-2.73 \substack{+0.31 \\ -0.35}$ & 
    $-2.66 \substack{+0.23 \\ -0.28}$ & 
    $-2.61 \substack{+0.17 \\ -0.23}$\\
    $\log \hspace{0.04cm} $\ce{CO2} & -3.44 & 
    $-4.52 \substack{+2.19 \\ -3.49}$ & 
    $-4.40 \substack{+2.11 \\ -3.69}$ & 
    $-4.71 \substack{+1.81 \\ -3.56}$ & 
    $-4.00 \substack{+1.09 \\ -3.61}$\\
    $\log \hspace{0.04cm} $\ce{CH4} & -5.77 & 
    $-6.64 \substack{+1.71 \\ -2.10}$ & 
    $-6.96 \substack{+1.81 \\ -2.02}$ & 
    $-7.28 \substack{+1.72 \\ -1.85}$ &  
    $-7.19 \substack{+1.62 \\ -1.89}$\\
    $\log \hspace{0.04cm} $\ce{N2O} & -6.55 & 
    $-4.41 \substack{+2.16 \\ -3.55}$ & 
    $-5.90 \substack{+2.44 \\ -2.73}$ & 
    $-6.55 \substack{+2.21 \\ -2.34}$ & 
    $-6.59 \substack{+2.06 \\ -2.28}$\\
    ${\mathrm{\log \hspace{0.04cm} P_0}}$ & 0.0 & 
    $0.61 \substack{+0.34 \\ -0.39}$ & 
    $0.40 \substack{+0.27 \\ -0.26}$ & 
    $0.39 \substack{+0.21 \\ -0.23}$ & 
    $0.38 \substack{+0.19 \\ -0.19}$\\
    $\mathrm{R_p}$ & 1.007 & 
    $0.76 \substack{+0.12 \\ -0.07}$ & 
    $0.86 \substack{+0.10 \\ -0.08}$ & 
    $0.88 \substack{+0.08 \\ -0.07}$ & 
    $0.90 \substack{+0.06 \\ -0.05}$\\
    $\mathrm{g}$ & 9.66 & 
    $13.07 \substack{+7.57 \\ -7.14}$ & 
    $13.51 \substack{+7.31 \\ -6.52}$ & 
    $13.67 \substack{+6.92 \\ -5.89}$ & 
    $13.42 \substack{+6.73 \\ -4.82}$\\
    $\mathrm{T}$ & 289 & 
    $321 \substack{+96 \\ -97}$ & 
    $321 \substack{+72 \\ -63}$ & 
    $310 \substack{+47 \\ -42}$ & 
    $305 \substack{+36 \\ -32}$ \\
    $\mathrm{\lambda_1}$ & 0.72 & 
    $0.723 \substack{+0.015 \\ -0.014}$ & 
    $0.723 \substack{+0.008 \\ -0.007}$ & 
    $0.723 \substack{+0.005 \\ -0.005}$ & 
    $0.723 \substack{+0.004 \\ -0.004}$\\
    $\mathrm{\lambda_2}$ & 1.40 & 
    $1.86 \substack{+0.46 \\ -0.48}$ & 
    $1.42 \substack{+0.33 \\ -0.11}$ & 
    $1.37 \substack{+0.08 \\ -0.06}$ & 
    $1.36 \substack{+0.07 \\ -0.04}$\\
    $\mathrm{A_{s,1}}$ & 0.09 & 
    $0.104 \substack{+0.027 \\ -0.024}$ & 
    $0.101 \substack{+0.012 \\ -0.012}$ & 
    $0.100 \substack{+0.009 \\ -0.009}$ & 
    $0.099 \substack{+0.007 \\ -0.007}$\\
    $\mathrm{A_{s,2}}$ & 0.15 & 
    $0.25 \substack{+0.05 \\ -0.06}$ & 
    $0.21 \substack{+0.04 \\ -0.04}$ & 
    $0.20 \substack{+0.03 \\ -0.03}$ & 
    $0.19 \substack{+0.02 \\ -0.02}$\\
    $\mathrm{A_{s,3}}$ & 0.06 & 
    $0.19 \substack{+0.25 \\ -0.11}$ & 
    $0.11 \substack{+0.05 \\ -0.03}$ & 
    $0.10 \substack{+0.03 \\ -0.02}$ & 
    $0.10 \substack{+0.02 \\ -0.02}$\\ 
	\enddata
	\tablecomments{$S/N$ is defined at 0.55\,$\micron$ (see Section~\ref{subsubsec:noise}).}
\end{deluxetable}

\newpage

\subsection{Surface Properties}

Figure~\ref{fig:retrievedhistograms} (bottom) highlights trends in the retrieved surface albedo parameters. As discussed in Sections~\ref{sec:results1} and \ref{sec:results2}, our main results are: i) including wavelength dependent surface albedo in retrievals can improve the accuracy of atmospheric inferences; and ii) even for low to moderate signal-to-noise ratios one can constrain the wavelength-dependent surface albedo for an Earth-analog planet. We highlight here that the $\lambda_1$ posterior demonstrates that a sharp change occurs in the surface albedo around 0.72\,$\micron$, even for $(S/N)_{\rm{ref}} = 5$, which is consistent with the modern Earth's red edge. This albedo transition is remarkably well constrained, as indicated by the narrow 1$\sigma$ intervals in Table~\ref{tab:significances}. Similarly, the posteriors for $\lambda_2$ at  $(S/N)_{\rm{ref}} \geq 10$ indicate that there is another sharp feature in the wavelength-dependent surface albedo around 1.4\,$\micron$. Compared with the posteriors for the bulk planetary properties and molecular abundances, these results suggest that surface albedo changes are one of the most reliable features to detect in reflection spectra of Earth-like planets.

\newpage

\subsection{The Influence of Clouds}

\begin{figure}
    \centering
    \includegraphics[trim = 1.0cm 0.0cm 1.0cm 0.5cm, width=0.90\columnwidth]{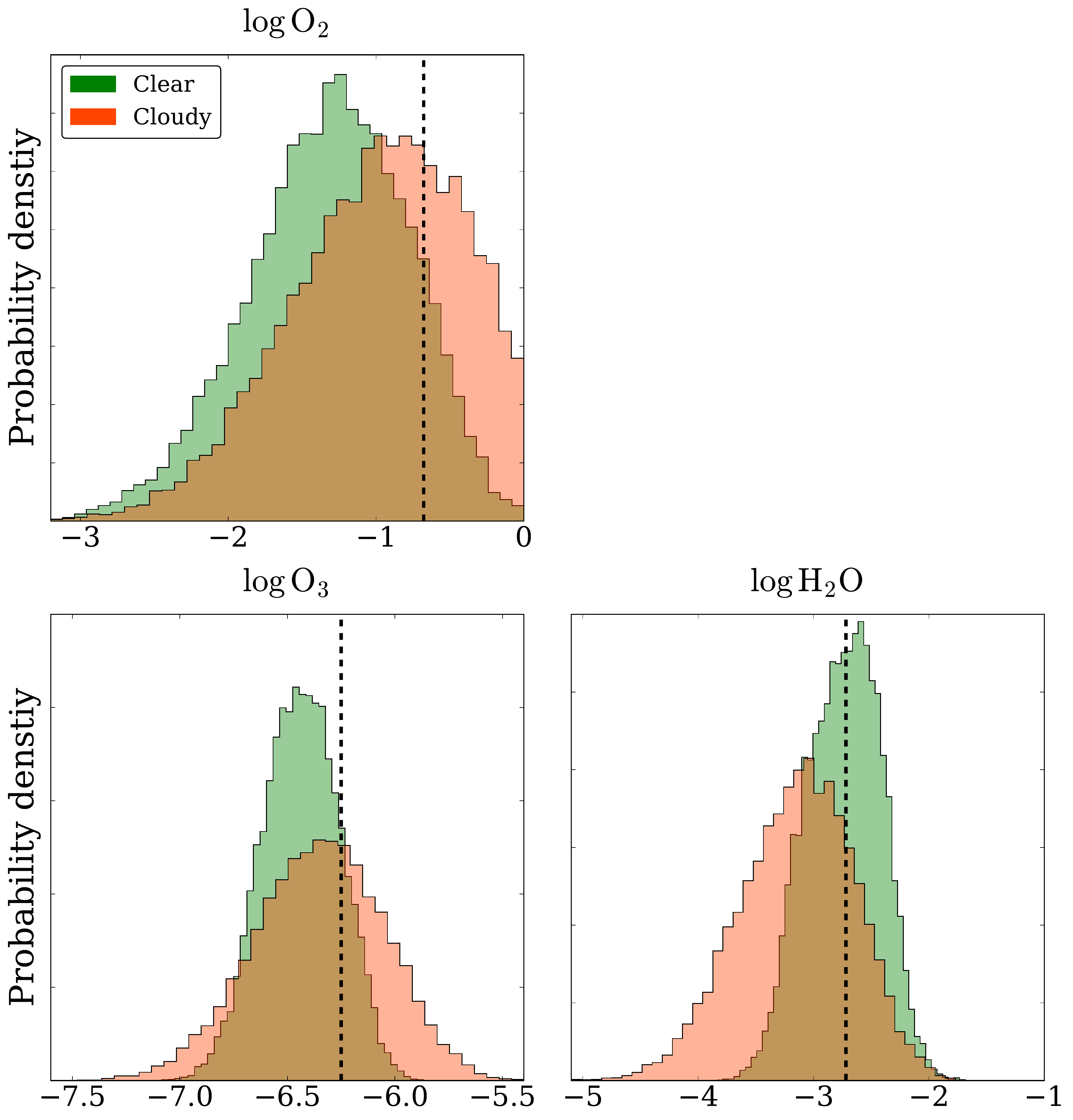}
    \caption{Retrieved molecular abundances from reflection spectra of an Earth-like exoplanet with a clear (green) and cloudy (orange) atmosphere. The simulated data used has $R = 70$ and $(S/N)_{\rm{ref}} = 10$). The ground truth reference values are overlaid (black lines, see Table~\ref{tab:parameters}).}
    \label{fig:molecular_histograms_cloud_comparison}
\end{figure}

While Figure~\ref{fig:retrievedhistograms} corresponded to cloud-free models, clouds can also increase the uncertainty on the retrieved atmospheric composition. Figure~\ref{fig:molecular_histograms_cloud_comparison} shows that a cloud deck increases the abundance uncertainties for detectable species. The cloud parameters broaden the 1$\sigma$ constraint for the oxygen, ozone, and water vapor abundances (from 0.55\,dex to 0.61\,dex for \ce{O2}; from 0.18\,dex to 0.30\,dex for \ce{O3}, and from 0.33\,dex to 0.50\,dex for \ce{H2O}). These effects can be attributed to the degeneracy that emerges between the location of the cloud-base and the gas mixing ratios (see Appendix~\ref{Appendix_C}). Despite the broader distributions, the \ce{O2, O3, and H2O} abundances are still retrieved to within 1$\sigma$ of their reference values when clouds are included in our model.

\begin{figure}
    \centering
    \includegraphics[width=0.90\columnwidth]{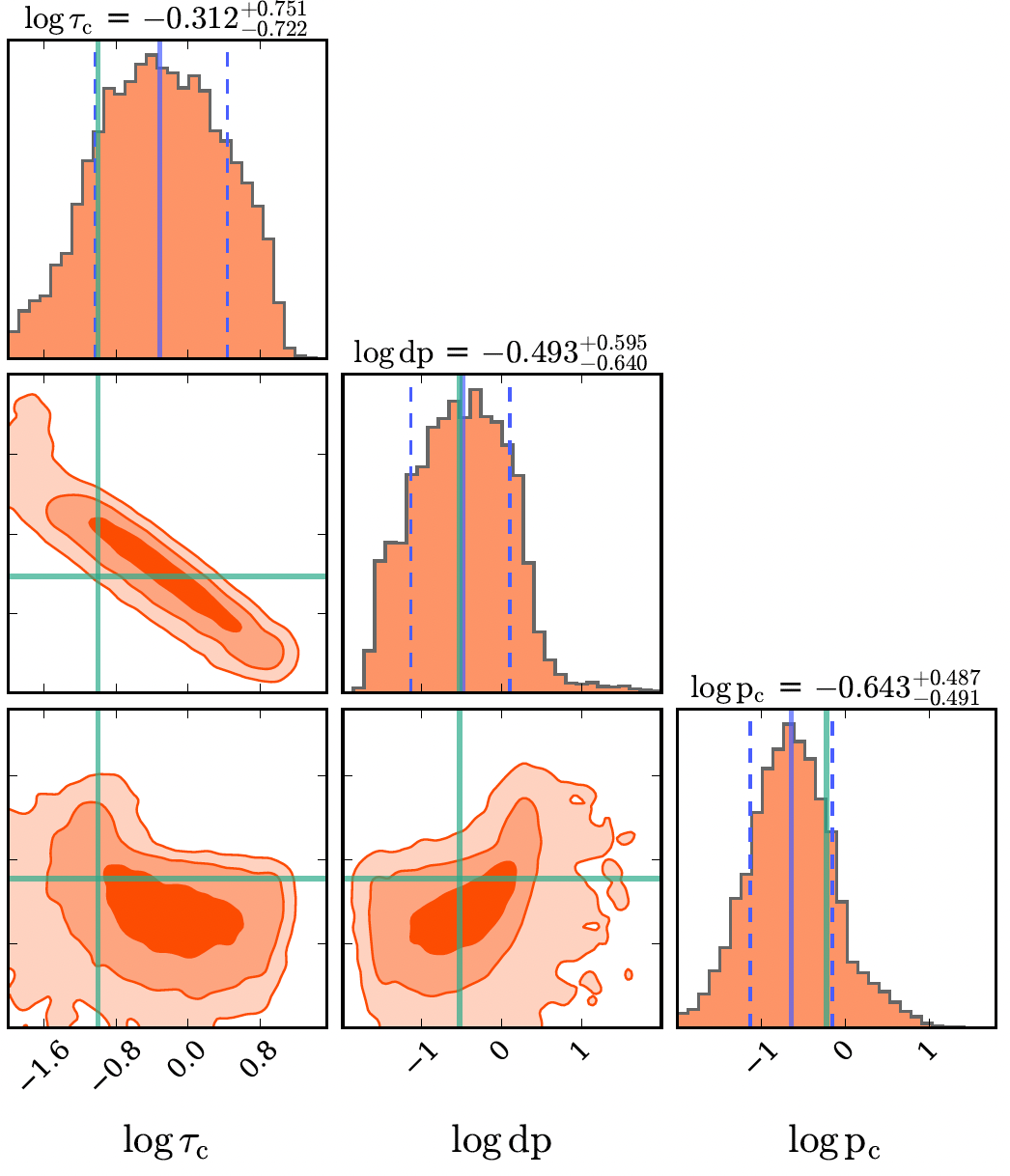}
    \caption{Retrieved cloud parameters for simulated data at $R = 70$ and $(S/N)_{\rm{ref}} = 10$. The retrieved median values (blue solid lines) agree with the reference values (green lines) within the 1$\sigma$ confidence regions (dashed lines).}
    \label{fig:cloud_params}
\end{figure}

Finally, Figure~\ref{fig:cloud_params} shows the cloud parameter constraints for $R = 70$ and $(S/N)_{\rm{ref}} = 10$ (corresponding to the retrieved spectrum in Figure~\ref{fig:retrievedcloudyspectra}). The retrieved optical depth ($\log \tau$), pressure extent ($\log dp$), and base pressure ($\log p_c$) of the cloud deck are all correctly retrieved within 1$\sigma$. The posteriors for $\log \tau$ and $\log dp$ are broad due to the degenerate nature of these parameters. However, the bounded constraint on $\log p_c$ demonstrates that our retrieval technique correctly identifies the presence of a cloud deck as a necessary model component distinct from the wavelength-dependent surface albedo.

\section{Summary and Discussion} 
\label{sec:discussion}

In this study, we investigated the potential to retrieve wavelength-dependent surface properties from reflection spectra of rocky exoplanets and the influence of such surfaces on the retrieval of molecular abundances and bulk planetary properties. We accomplished this by developing and implementing a Bayesian retrieval approach to infer wavelength-dependent surface properties from simulated observations of a self-consistent modern Earth-like planet. Our retrieval analysis demonstrated that it is possible to infer wavelength-dependent surface information at moderate signal-to-noise ratios. Our key results from this work include the following:

\begin{enumerate}
    \item Reflected-light retrievals of terrestrial exoplanets should account for wavelength-dependent surface albedos to achieve unbiased results. If one assumes a constant surface albedo, some optical and near-infrared data can be poorly fit and the retrieved mixing ratios can be biased.
    \item We introduced a five-parameter prescription to account for sharp `edge-like' changes in the surface albedo at \emph{a priori} unknown wavelengths. We demonstrated that this parameterization can recover realistic surface albedo profiles from reflection spectra of a modern Earth analog.
    \item Wavelength-dependent surface albedos can be readily retrieved from cloud-free reflection spectra. Even with $S/N = 5$, one can infer evidence of Earth-like wavelength-dependent surface features (2.7$\sigma$). The easiest feature to detect for our modern Earth analog is a sharp change in the surface albedo at visible wavelengths (i.e. the modern Earth's red edge). Improved data quality ($S/N \geq 10$) allows a secondary albedo change in the infrared to be constrained as well.
    \item Clouds can lower the detection significance of wavelength-dependent surface features. Nevertheless, one can still infer modern Earth-like surface features in the presence of clouds at $S/N = 10$ (2.9$\sigma$). Also, the wavelength of the modern Earth's red edge can still be reliably retrieved.
    \item Higher spectral resolution improves constraints on wavelength-dependent surface features for cloud-free modern Earth-like models. Specifically, data at $R = 140$ outperforms $R = 70$ data in identifying surface features at infrared wavelengths.
    \item The mixing ratios of gases with strong absorption features in the optical and near-infrared --- namely \ce{O3}, \ce{O2}, and \ce{H2O} --- can be precisely retrieved ($<$ 0.6\,dex) with moderate quality data ($S/N = 10$) when wavelength dependent surface albedos are included in retrievals. Gases with weaker infrared absorption, such as \ce{CH4}, \ce{CO2}, and \ce{N2O}, are largely unconstrained by reflection spectra (upper limits only).
    \item Several other planetary properties (e.g. planet radius, temperature, and cloud properties) can be retrieved from reflection spectra. However, the planetary gravity can not be determined from reflection spectra alone.
    
\end{enumerate}

We proceed to discuss the implications of our findings.

\subsection{Retrievability of Diverse Surface Compositions}

Wavelength-dependent surface albedos significantly impact the reflection spectra of directly-imaged terrestrial exoplanets (e.g. Figure~\ref{fig:spectra}). Our retrieval analysis demonstrates that future direct imaging missions could find evidence of wavelength-dependent surface features and constrain the shape of surface albedo profiles. Our results complement and expand on the recent study by \citet{Wang2022}, by offering a novel parameterization to retrieve changes in the surface albedo at \emph{a priori} unknown wavelengths. In particular, our demonstration that sharp features like the modern Earth's vegetation red edge can be reliably retrieved (to 150\,nm for cloud-free models at $S/N = 5$) is very promising for a proposed future large IR/Optical/UV space-based telescope \citep[e.g.][]{NationalAcademiesofSciences2021}. While we find that clouds can result in overestimated surface albedos (in agreement with results from \citealt{Feng2018}, \citealt{Robinson2022}, and \citealt{Damiano2022}), the red edge's wavelength location can nonetheless be correctly retrieved for modern Earth-like planets even for Earth-like cloud coverage.

\begin{figure}[t!]
    \centering
    \includegraphics[trim = 0.5cm 0.2cm 0.1cm 0.0 cm, width=\columnwidth]{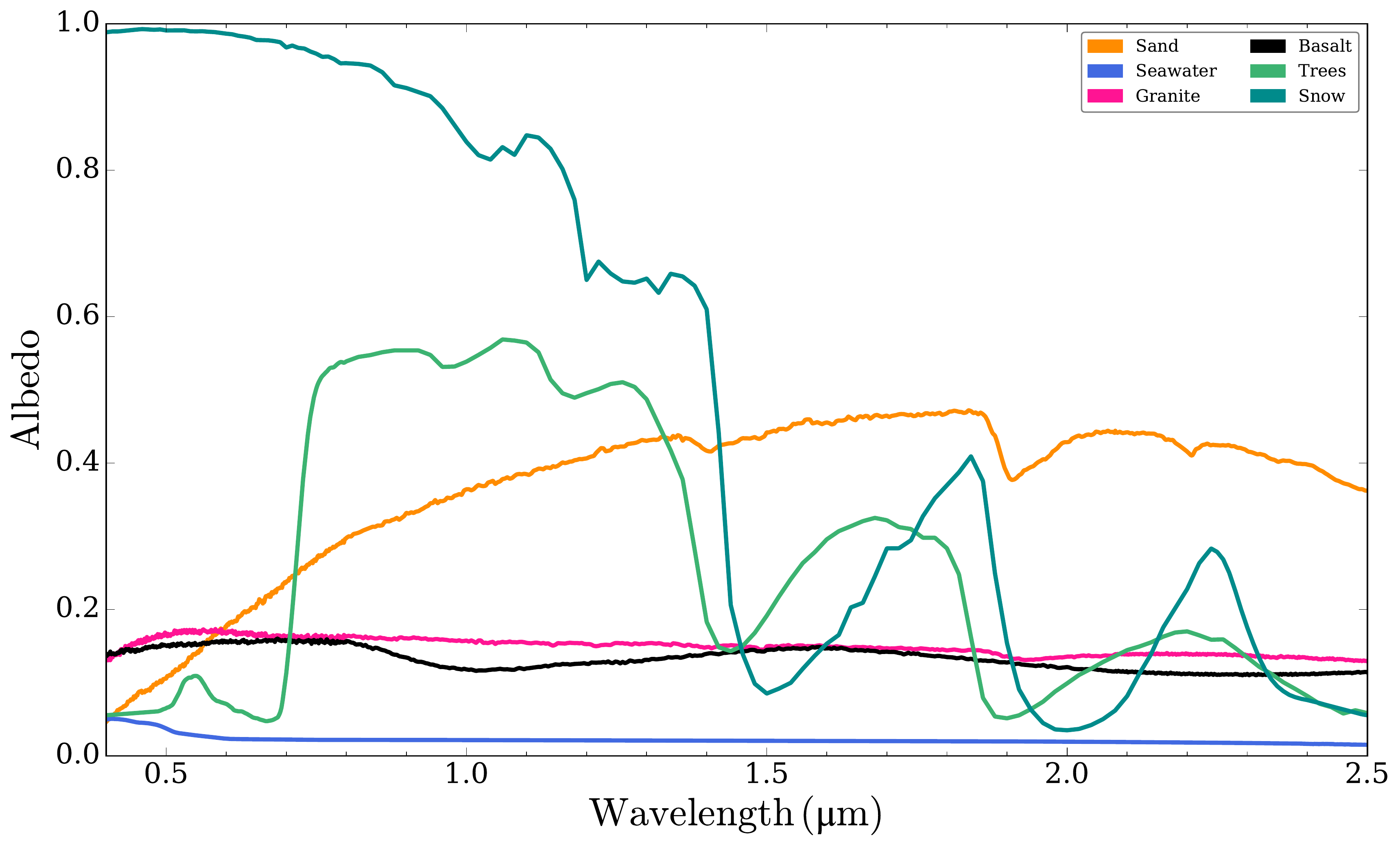}
    \caption{Wavelength-dependent albedos of materials commonly found on Earth's surface.}
    \label{fig:albedos_comparison}
\end{figure}

More generally, detecting wavelength-dependent surface albedos from reflection spectra offers the opportunity to constrain the surface composition of rocky exoplanets. Other materials like sand, basalt, and granite, which cover substantial regions of Earth's surface, have unique albedo profiles (see Figure~\ref{fig:albedos_comparison}) and shape the reflection spectra of Earth-like exoplanets \citep[see e.g.][]{Madden20}. These profiles could potentially be extracted from spectra of rocky worlds whose surfaces are dominated by these materials \citep{Pham2021,Pham2022}. However, for retrieval purposes the flexibility of our surface albedo parameterization (Equation~\ref{eq:3}) would need to be tested for these surface compositions. Our parameterization was inspired by the modern Earth's surface albedo, hence it is able to locate sharp albedo changes such as the red edge. Future work should investigate the flexibility of our parameterization for other surfaces, such as oceans or deserts, to determine if a generalized parameterization is necessary.

\begin{figure*}[t!]
    \centering
    \includegraphics[trim={0 -0.33cm 0 1.0cm}, width=0.962\columnwidth]{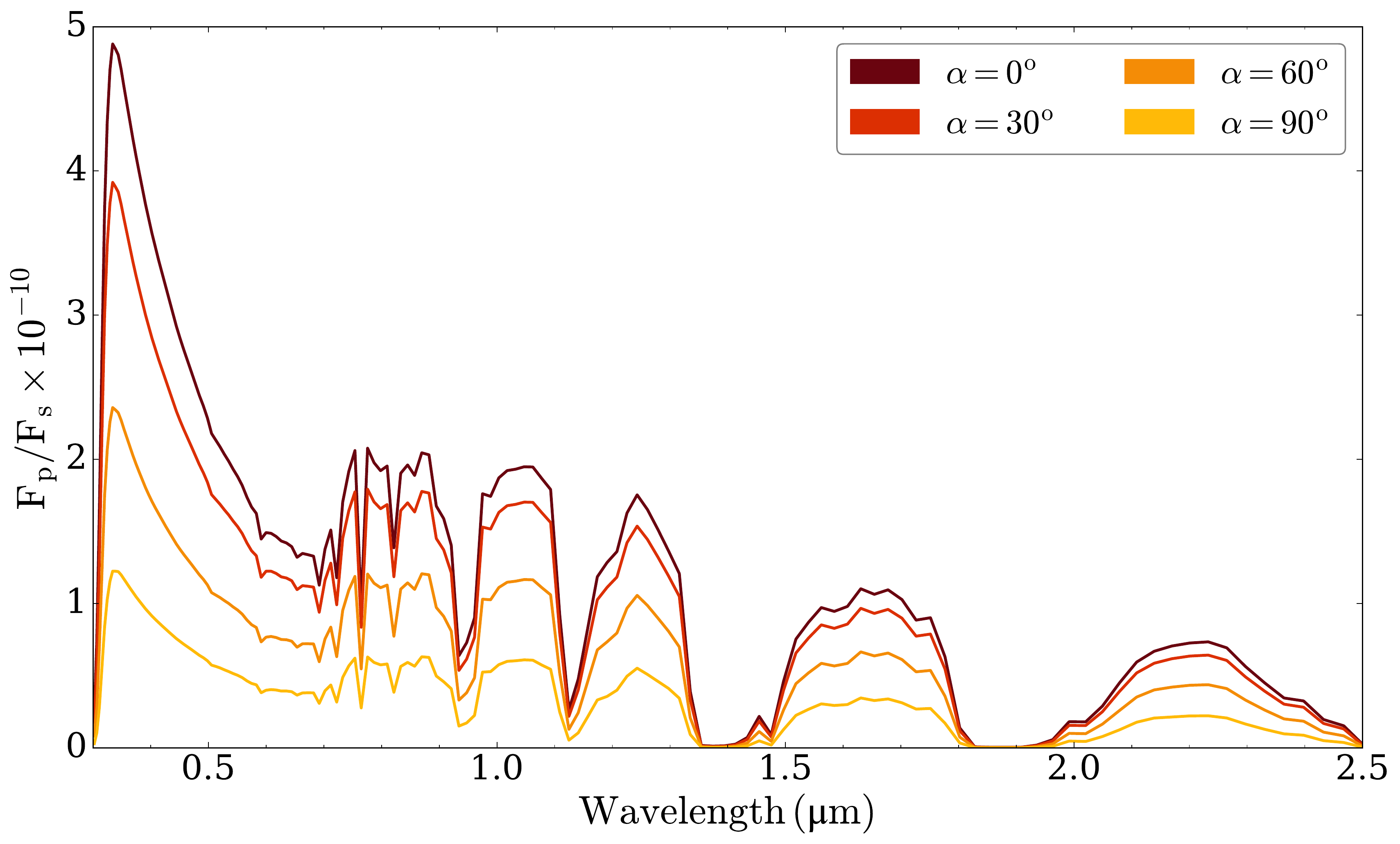}
    \includegraphics[width=0.962\columnwidth]{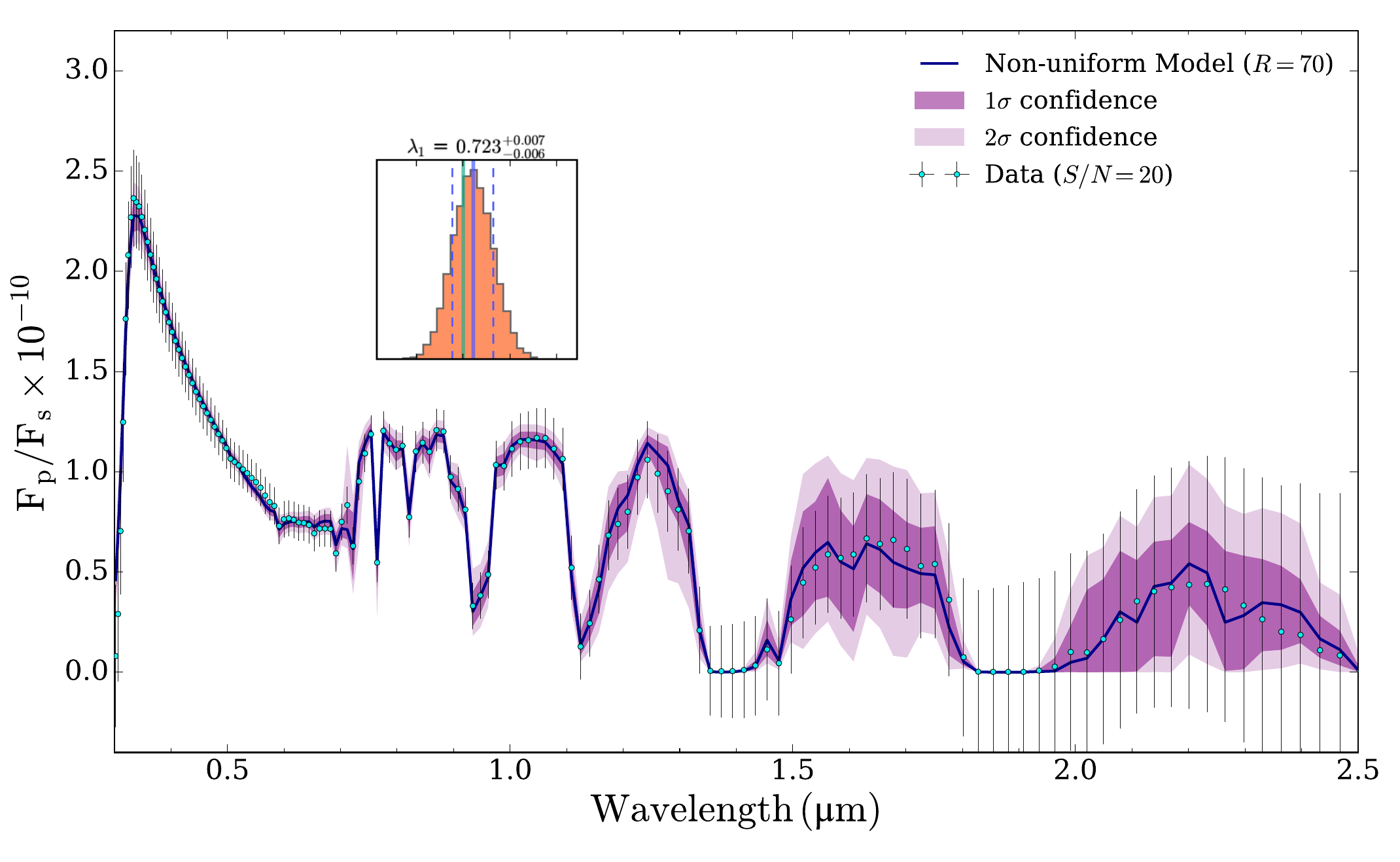}
    \caption{\textit{Left:} Impact of different orbital phase angles on reflection spectra of an Earth-like exoplanet. The sequence of models have the same atmospheric and surface properties as Figure~\ref{fig:spectra}, but with partial illumination. \textit{Right:} retrieved reflection spectrum of a cloud-free Earth-like exoplanet at orbital phase of $60^{\circ}$ with a wavelength-dependent surface albedo (for simulated data at $R = 70$ and $(S/N)_{\rm{ref}} = 20$). The inset shows the posterior distribution of the retrieved red edge wavelength.}
    \label{fig:phases}
\end{figure*}

An additional caveat is that our retrievals have focused on zero orbital phase (i.e. full illumination), while future direct imaging observations will be constrained to higher phase angle. The left panel of Figure \ref{fig:phases} shows the impact of higher phase angles through PICASO calculations at orbital phases of $0^{\circ}$, $30^{\circ}$, $60^{\circ}$, and $90^{\circ}$ for the model in Figure~\ref{fig:spectra}. We see that orbital phase acts to scale down $F_p/F_*$, while preserving the relative amplitude of the red edge relative to the surrounding continuum. The right panel of Figure \ref{fig:phases} shows the impact of a non-zero orbital phase on a retrieval of our cloud-free Earth-like exoplanet model. Due to the computational requirements of non-zero orbital phase retrievals, for this demonstration we consider only an orbital phase of $60^{\circ}$ and observations with $(S/N)_{\rm{ref}} = 20$ and $R = 70$. We see that the location of the vegetation red edge can still be constrained for non-zero phases, but the uncertainty becomes about $50\%$ larger due to the lower flux ratio. Future retrieval studies could investigate the impact of a wide range of non-zero orbital phases \citep[e.g.][]{Nayak2017}, or multi-phase observations \citep{Damiano2020b, Carrion-Gonzalez2021}, on wavelength-dependent surface albedo constraints. Nevertheless, our results show great promise for the detectability and characterization of rocky exoplanet surfaces from reflection spectra.

\subsection{Opportunities for Detecting Surface Biosignatures}

The potential habitability of Earth-like exoplanets orbiting Sun-like stars could be assessed by constraining wavelength-dependent surface properties. Future direct-imaging missions will focus on characterizing these atmospheres to search for biosignature gas pairs like \ce{O3} combined with \ce{CH4}. Our results suggest that such missions can also retrieve surface spectral features, and that not accounting for wavelength-dependent surfaces can bias the retrieved abundances of biosignature gas pairs. We stress that including wavelength-dependent surface albedos in retrievals is an \emph{opportunity}, since it enables an expanded mission science case including searches for surface biosignatures. The vegetation red edge is one candidate, but its universality remains uncertain. Exoplanets could have reflectance edges at different wavelengths \citep[see e.g.][]{Kiang2007} or photosynthetic organisms that do not show red edge features  \citep[see e.g.][]{Cockell2009}. Some minerals also exhibit sharp spectral features near optical wavelengths (e.g., \citealt{Seager2005,O'Malley-James2018}). Thus, any detection of reflectance edges would need to be carefully placed in context with other signatures of habitability before attributing a biological origin.

The promising detectability of the modern Earth's red edge also suggests that other surface biosignatures would benefit from retrieval studies. One such alternative surface biosignature is biofluorescence. On Earth, coral and other organisms absorb harmful shortwave radiation and re-emit it at longer wavelengths as a protection mechanism. Like the vegetation red edge, biofluorescence can dramatically increase a planet's brightness at specific wavelengths \citep{O'Malley-James2019, O'Malley-James2018b}. Biofluorescence could therefore manifest as a time-dependent spectral edge that may be retrievable from reflection spectra of Earth-like exoplanets.

\subsection{Constraining Atmospheric Properties of Earth-like Exoplanets via Reflected Light}

One of the primary science goals of future space-based observatories with direct imaging capabilities will be to characterize the atmospheres of Earth-like exoplanets orbiting Sun-like stars. Our results indicate that a large space-based observatory with the capability to achieve $F_p / F_* \sim 10^{-10}$ could precisely constrain the abundances of several biosignature pair gases on modern Earth analogs. Assuming a cloud-free atmosphere, we showed that the \ce{O2}, \ce{O3}, and \ce{H2O} abundances can be constrained within 0.6\,dex for $S/N = 10$ and $R = 70$. Crucially, the \ce{O3} abundance can be constrained to within a factor of 2 (0.3\,dex) even for $S/N = 5$. However, it will be more challenging for these missions to detect biosignature gases with weak absorption features. For instance, we could only place an upper limit on the \ce{CH4} mixing ratio for a modern Earth-like analog. Furthermore, clouds can also impact our ability to constrain and detect some biosignatures. Clouds broaden the uncertainties in the retrieved molecular abundances of \ce{O2}, \ce{O3}, and \ce{H2O}, which makes it more challenging to constrain their abundances.

Our retrieval results suggest that meaningful information can be extracted from reflected light spectra of a modern Earth analog, even with a lower SNR than indicated by previous work, if the available data covers an expanded wavelength range. \citet{Feng2018} found that $S/N = 15$ is generally a prerequisite to constrain the abundances of \ce{O2}, \ce{O3}, and \ce{H2O}. In comparison, we find that these gases can be precisely constrained for $S/N \geq 10$. These differences are mainly attributable to the wavelength range of the simulated data --- we use 0.3--2.5\,$\micron$, while \citet{Feng2018} considered 0.4--1.0\,$\micron$. The longer wavelength coverage in our retrievals decreased the minimum $S/N$ necessary to constrain \ce{H2O} because of the three additional water features at 1.1, 1.4, and 1.9\,$\micron$. Similarly, the short wavelength coverage of \ce{O3} absorption lowered the $S/N$ necessary for constraining the \ce{O3} abundance. However, the difference between the minimum signal-to-noise ratios for retrieving the \ce{O2} abundance are: i) due to our differing noise models (we assumed an agnostic detector efficiency while \citet{Feng2018} based their model on Roman Space Telescope-like detectors); and ii) due to the wavelength dependent surface features we added to the retrieval process.

\subsection{Characterizing Exoplanets Orbiting Sun-like Stars}

Ultimately, a key driver of exoplanet science is the characterization of potentially habitable planets around Sun-like stars. The Astro 2020 Decadal Survey \citep{NationalAcademiesofSciences2021} specifically highlights the goal of searching for atmospheric biosignatures on Earth-like exoplanets orbiting Sun-like stars. Our findings indicate that a future direct imaging mission observing reflected light could also detect surface biosignatures --- including the vegetation red edge --- via the retrieval of a wavelength-dependent surface albedo. Our odds of detecting life in the solar neighborhood can only be enhanced by considering all the ways life shapes its host planet. For a spectral edge encoded in light from a distant star may, one day, illuminate the surface of a world not too dissimilar to our own.

\begin{acknowledgments}

We thank Natasha Batalha and Zifan Lin for helpful discussions. We also thank the anonymous referee for a helpful report. J.G.B. is supported by Cornell University's Ronald E. McNair Post-baccalaureate Achievement Program and the Carl Sagan Institute.

\end{acknowledgments}

\software{PICASO \citep{Batalha2019}}, coronagraph \citep{Lustig2019coronagraph}

\section*{Data Availability}
The realistic Earth-like surface albedo and the raw albedo files used in this work are available at \url{https://doi.org/10.5281/zenodo.6977238}.
\\

\appendix

\section{Retrievals with Gaussian Scatter} \label{Appendix_A}

\begin{figure*}
    \centering
    \includegraphics[width=0.85\textwidth]{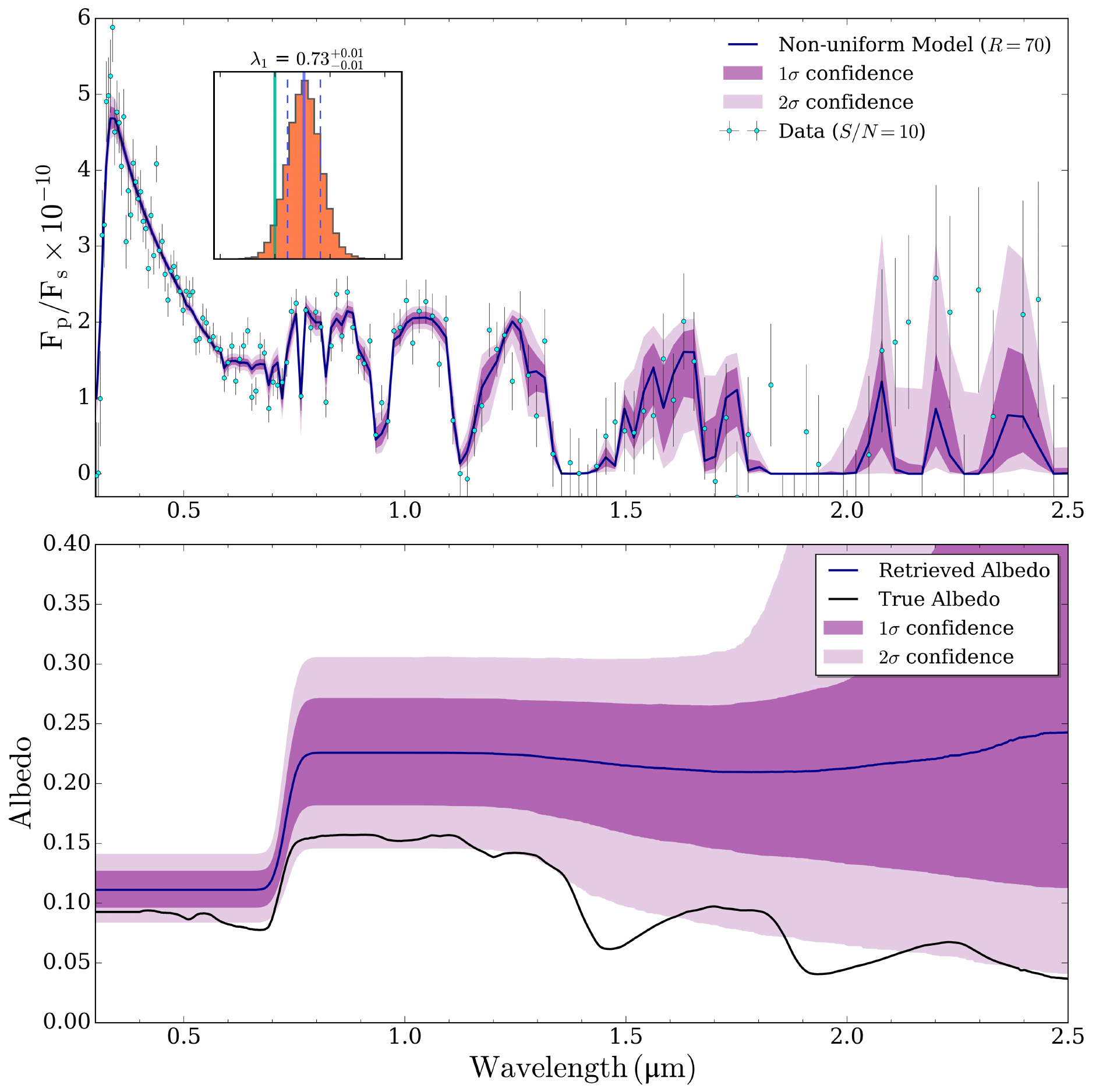}
    \caption{\textit{Top:} retrieved reflection spectrum of a cloud-free Earth-like exoplanet with a wavelength-dependent surface albedo (for simulated data with gaussian scatter at $R = 70$ and $(S/N)_{\rm{ref}} = 10$). The inset shows the posterior distribution of the retrieved red edge wavelength. \textit{Bottom:} The corresponding retrieved surface albedo profile (purple contours) compared to the true surface albedo (black line).}
    \label{fig:retrieval_gaussian}
\end{figure*}

Here, we show a retrieved reflection spectrum from synthetic observations with Gaussian scatter. Figure~\ref{fig:retrieval_gaussian} shows that the retrieved spectra captures the overall spectral morphology of the input model and that the location of the vegetation red edge is still constrained.

\section{Retrieval Model Validation} \label{Appendix_B}

\begin{figure*}
    \centering
    \includegraphics[width=0.85\textwidth]{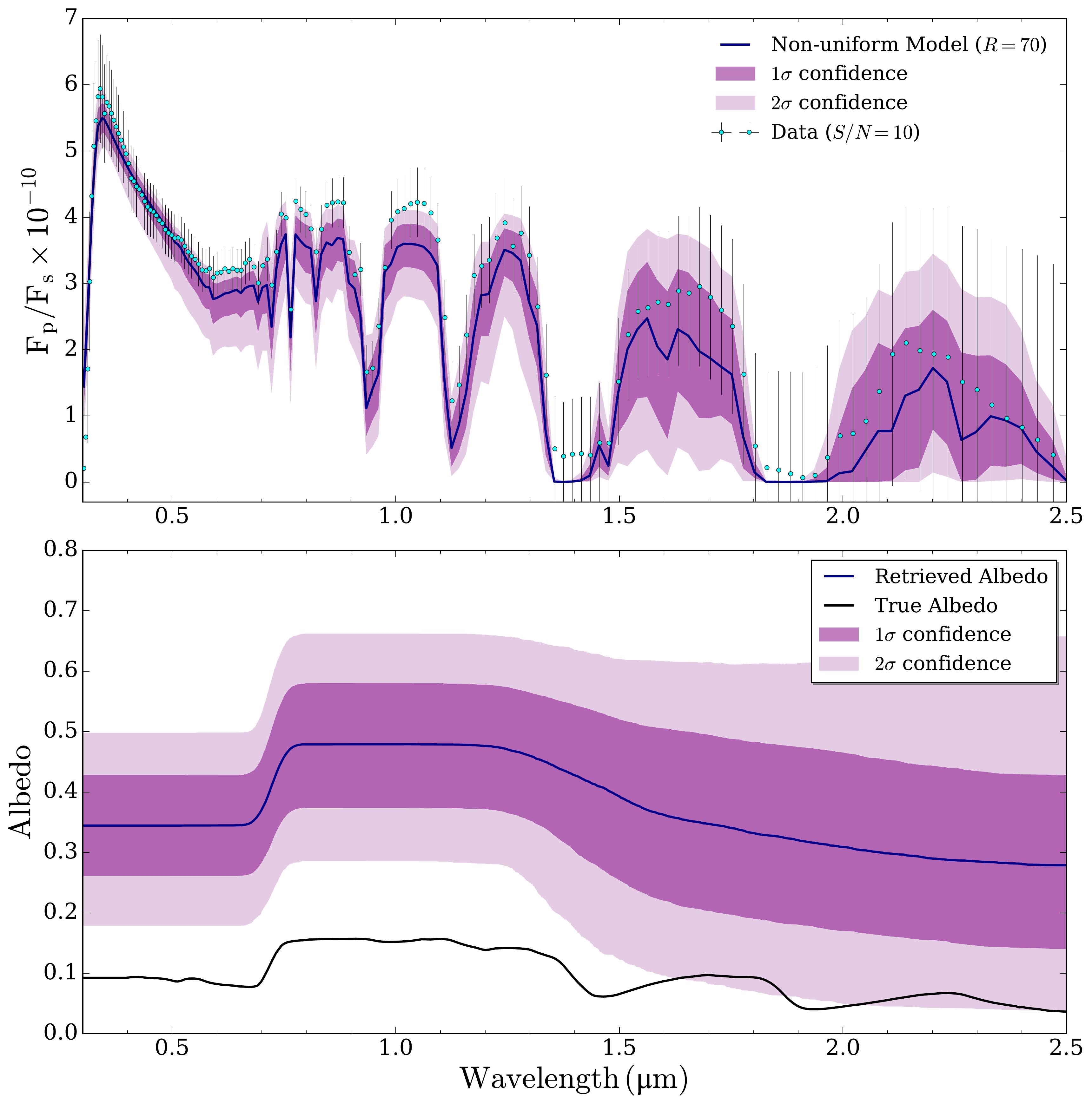}
    \caption{\textit{Top:} retrieved reflection spectrum of a cloudy Earth-like exoplanet with a wavelength-dependent surface albedo (for simulated data of the \citet{Robinson2011} model at $R = 70$ and $(S/N)_{\rm{ref}} = 10$). \textit{Bottom:} The corresponding retrieved surface albedo profile (purple contours) compared to a realistic Earth-like surface albedo (black line).}
    \label{fig:robinson_retrieval}
\end{figure*}

Here, we validate our retrieval framework using the model Earth spectrum from \citet{Robinson2011}, which has been validated against Earthshine data. The simulated data was generated by binning the model spectrum from its native resolution to R=70. We then used our noise model to simulate data at $(S/N)_{\rm{ref}} = 10$. Figure \ref{fig:robinson_retrieval} shows that our retrieval model reproduces the general spectral morphology of the \citet{Robinson2011} model. As with the cloudy 1D simulated data, the retrieved albedo is brighter than the model surface albedo. Nevertheless, our retrieval results indicate that the location of the red edge can still be constrained at $(S/N)_{\rm{ref}} = 10$.

\section{Full Posterior Distributions} \label{Appendix_C}

\begin{figure*}
    \centering
    \includegraphics[width=\textwidth]{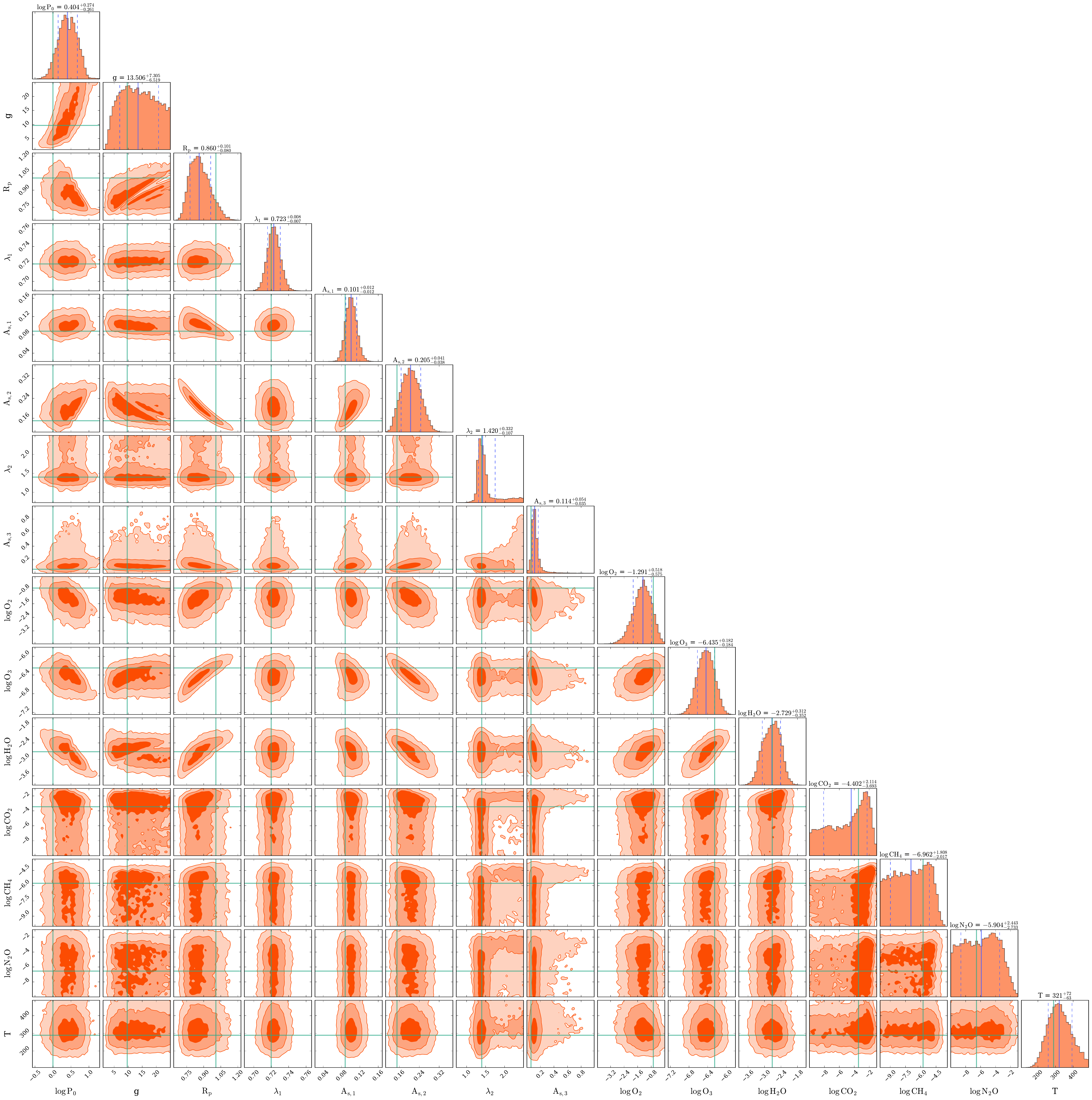}
    \caption{The full posterior distribution for the cloud-free simulated data at $R = 70$ and $(S/N)_{\rm{ref}} = 10$. The retrieved median (solid blue lines) and 16\textsuperscript{th} and 84\textsuperscript{th} percentiles (dashed blue lines) for each parameter are overlaid for comparison with the ground truth reference values (green lines). Note that the mixing ratio reference values represent the average value from the surface to 25\,km altitude. Similarly, the surface albedo parameter reference values represent the average albedo over a wavelength range.}
    \label{fig:clear_corner}
\end{figure*}

\begin{figure*}
    \centering
    \includegraphics[width=\textwidth]{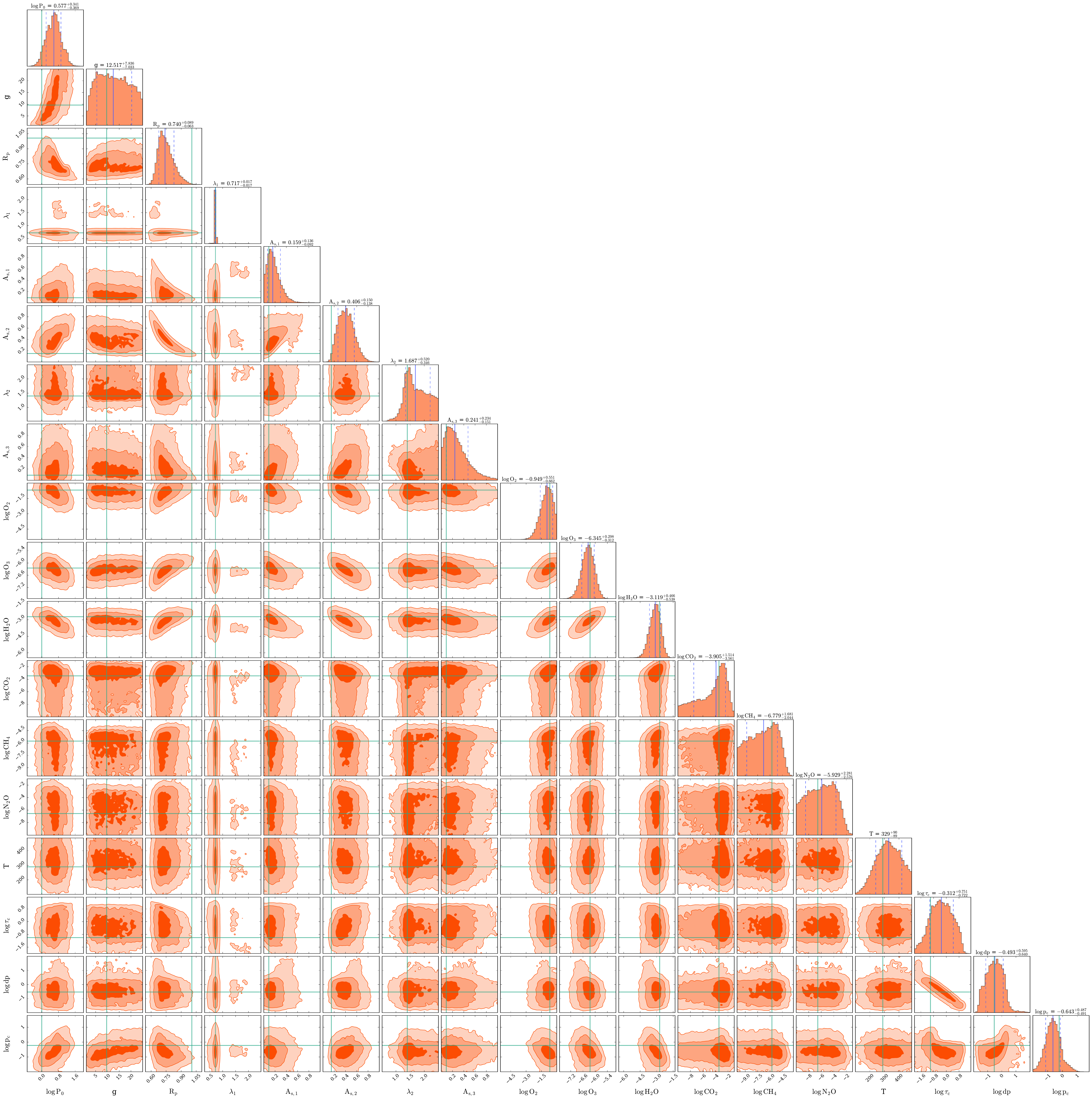}
    \caption{The full posterior distribution for the cloudy simulated data at $R = 70$ and $(S/N)_{\rm{ref}} = 10$. The retrieved median (solid blue lines) and 16\textsuperscript{th} and 84\textsuperscript{th} percentiles (dashed blue lines) for each parameter are overlaid for comparison with the ground truth reference values (green lines).}
    \label{fig:cloudy_corner}
\end{figure*}

Here, we include full posterior distributions from representative retrievals of our modern Earth analog. Figure~\ref{fig:clear_corner} shows the posterior distribution for a cloud-free retrieval at $(S/N)_{\rm{ref}} = 10$ and $R = 70$ including a wavelength-dependent surface albedo. Similarly, Figure~\ref{fig:cloudy_corner} shows the posterior for the cloudy scenario discussed in Section~\ref{subsec:cloud_sensitivity}.

\bibliography{red_edge}{}
\bibliographystyle{aasjournal}

\end{document}